\documentclass[twocolumn,showpacs,amssymb,nobibnotes,aps,floats,psfig,prb]{revtex4-1}
\usepackage{textcomp,amssymb,graphicx,epsf}
\usepackage{hyperref}
\usepackage{epstopdf}
\usepackage{amsbsy}
\usepackage{amsmath}
\usepackage{mathtools}
\usepackage {xcolor}
\usepackage{graphicx,epsfig}
\usepackage{float}
\begin{document}
\title{
Study of the ferromagnetic-insulator phase in manganites}
\author{Sanjukta Paul}
\author{Sudhakar Yarlagadda}
\affiliation{
CMP Div., Saha Institute of Nuclear Physics, HBNI,
Kolkata, India}
\date{\today}

\begin{abstract}
Understanding the coexistence of ferromagnetism and insulating behavior
in manganites is an unsolved problem.
We propose a localized-band model involving effective 
intermediate-range electron-electron (electron-hole) repulsion (attraction)
 generated by cooperative electron-phonon interaction. Double exchange mechanism,
 involving holes virtually hopping to nearest neighbors and back,
 produces magnetic polarons in an antiferromagnetic environment; 
 when these magnetic polarons  coalesce and percolate the system, we get a 
 ferromagnetic insulator. Ferromagnetism gets more pronounced when the holes (doping) increases
 or when the ratio  hopping/polaronic-energy dominates over superexchange-coupling/hopping.
\end{abstract}  


\maketitle

\pagebreak

\section{Introduction}
\label{intro}
Perovskite oxides, such as manganites, display a variety of 
orbital, charge, and spin orders when   the
parent oxide is doped. While significant progress has been made in characterizing
most of the phenomena in bulk doped materials, the understanding pertaining
to ferromagnetic insulator is still elusive.
The doped alloy ${\rm T_{1-x} D_x MnO_3}$ 
(where T refers to trivalent rare-earth elements such as La, Pr, Nd, etc.
and D refers to divalent alkaline elements Sr, Ca, etc.)
is an antiferromagnet when $x > 0.5$ with the nature of the antiferromagnet 
(i.e., A-, C-, CE-, or G-type antiferromagnet)
depending
on the compound and the dopant value  $x$ \cite{cnr,khomskii0,hotta}. Contrastingly, for $x < 0.5$, ${\rm T_{1-x} D_x MnO_3}$
is an intriguing  ferromagnetic insulator (FMI) at smaller values of x (i.e., $0.1 \lesssim x \lesssim 0.2$)
\cite{tokura,cheong,raveau} and is a ferromagnetic metal at higher dopings
in the manganite systems ${\rm La_{1-x} Sr_x MnO_3}$, ${\rm La_{1-x} Ca_x MnO_3}$, ${\rm Pr_{1-x} Sr_x MnO_3}$,
and ${\rm Nd_{1-x} Sr_x MnO_3}$.

For modeling the diverse orderings and for exploiting the 
functionality in these transition-metal oxides, one
needs effective Hamiltonians for various types of
interactions. 
Although the importance of strong electron-phonon interaction (EPI) has been pointed
much earlier \cite{littlewood} and significant  progress has been made some time ago in
numerically treating electron-phonon interaction in sizeable systems \cite{pinaki}, 
the treatment of cooperative EPI (involving quantum phonons)
was accomplished analytically only
more recently in two dimensions (2D) \cite{cbm2d}. 
It has been demonstrated
analytically in Ref. \onlinecite{cbm2d} that introducing cooperative effects, when 
 EPI is strong,  produces nearest-neighbor (NN), next-nearest-neighbor (NNN),
and next-to-next-nearest-neighbor (NNNN) interactions.
Furthermore, incorporating spin-spin interactions along with
 cooperative strong EPI is still an unsolved  analytic problem.

As regards  experiments pertaining to ferromagnetic-insulating regions,
while some suggested  microscopically homogeneous electronic properties \cite{dai1,dai2,jiang},
others speculate that coexistence of ferromagnetic metallic phases and antiferromagnetic
insulating phases leads to an inhomogeneous ferromagnetic insulating state \cite{markovich1,markovich2}.

  We will now argue, without considering any specific model, that ferromagnetic insulating phases are possible  at low doping
  in manganites
   by presenting below general theoretical points based on
    the essential features of manganites.
    
\noindent 1) Kinetic energy (KE) is quite small at low doping because bare hopping is small (caused by lower 
tolerance factor \cite{peter1}, cation disorder,  compatibility of distortions \cite{peter2}), and the
    electron-phonon coupling is strong.

\noindent 2)  Potential energy [from repulsive interactions, due to cooperative 
    EPI, that are intermediate-range, i.e., NN, NNN,
  NNNN, etc.] is much larger than KE; this leads to solid-type formation 
    with electrons being rendered essentially immobile and site localized. 
     The ground state is classical  and the state of the system can be expressed by a single 
    state in the occupation number basis with number density
    at each site either 1 or 0. 
    
    The fact that electrons are essentially site localized  also follows from the treatment
    in  Ref. \onlinecite{tvr}; then, only a localized polaronic band 
    is relevant and the upper wide band cannot overlap with the lower  narrow polaronic band.

    Furthermore,  a simple type of  phase separated state with ferromagnetic droplets (each containing one carrier)
    in an antiferromagnetic matrix was shown to be possible in Ref. \onlinecite{khomskii} . The mobility
    of these magnetic polarons is low and they are easily localized by disorder and Coulomb interactions.
    
    Thus the potential energy determines 
    the charge and spin order.

\noindent 3) Because of  cooperative strong EPI, 
    a NN electron-hole pair has a strong ferromagnetic interaction 
   [$t^2\cos^2(\theta/2)/(2E_{\rm JT})$
   with $E_{\rm JT}$ being the cooperative Jahn-Teller energy, $t$ the hopping term between the NN sites, and
   $\theta$ the angle between the NN core spins]. Hence, a robust ferromagnetic cluster is produced in the vicinity of a hole.

\noindent 4) Our model for magnetic interaction  applies to manganites with low density of localized
   holes.  In regions  away from the holes, the cooperative EPI retains essentially the same 
   orbital texture as in the undoped manganite. As a result,  in regions without holes, the magnetic interaction is A-AFM just
   as in the undoped manganite.
   As regards the regions with holes, since the holes are site  
   localized, the holes only virtually hop to a NN site and back and thus produce ferromagnetic coupling
   with NN electrons. This ferromagnetic coupling, between NN electron-hole pair, is much stronger than A-AFM coupling.

\noindent 5)  Presence of site localized holes  produces FMI clusters due to formation
    of magnetic polarons.  A hole will polarize NN electrons (and realistically speaking,
    NNN and NNNN electrons as well) through virtual hopping, thereby producing a magnetic 
    polaron. A collection of interacting magnetic polarons will produce a FMI region. 
    It is interesting to note that FMI regions are present at moderate doping in manganites that
    are narrow band (${\rm Pr_{1-x}Ca_x MnO_3}$), intermediate band (${\rm La_{1-x}Ca_x MnO_3}$)
    and wide band (${\rm La_{1-x}Sr_x MnO_3}$).

 Rest of the paper is organized as follows. In Sec. II, invoking cooperative electron-phonon-interaction physics,
 we obtain the effective Hamiltonian that is employed  to understand the FMI phase in manganites.
Next, in Sec. III, we outline our calculation procedure involving Monte Carlo technique, used to simulate
charge and spin configurations, and obtain the magnetization as a function of various parameters.
Then, in Sec. IV, we discuss our results obtained for systems with different hoppings at different
temperatures and dopings. Lastly, in Sec. V, we conclude and offer some perspectives.
 
\section{Effective Hamiltonian}
\label{mod_ham}
\begin{figure}[!htb]
 \includegraphics[width=9.0cm]{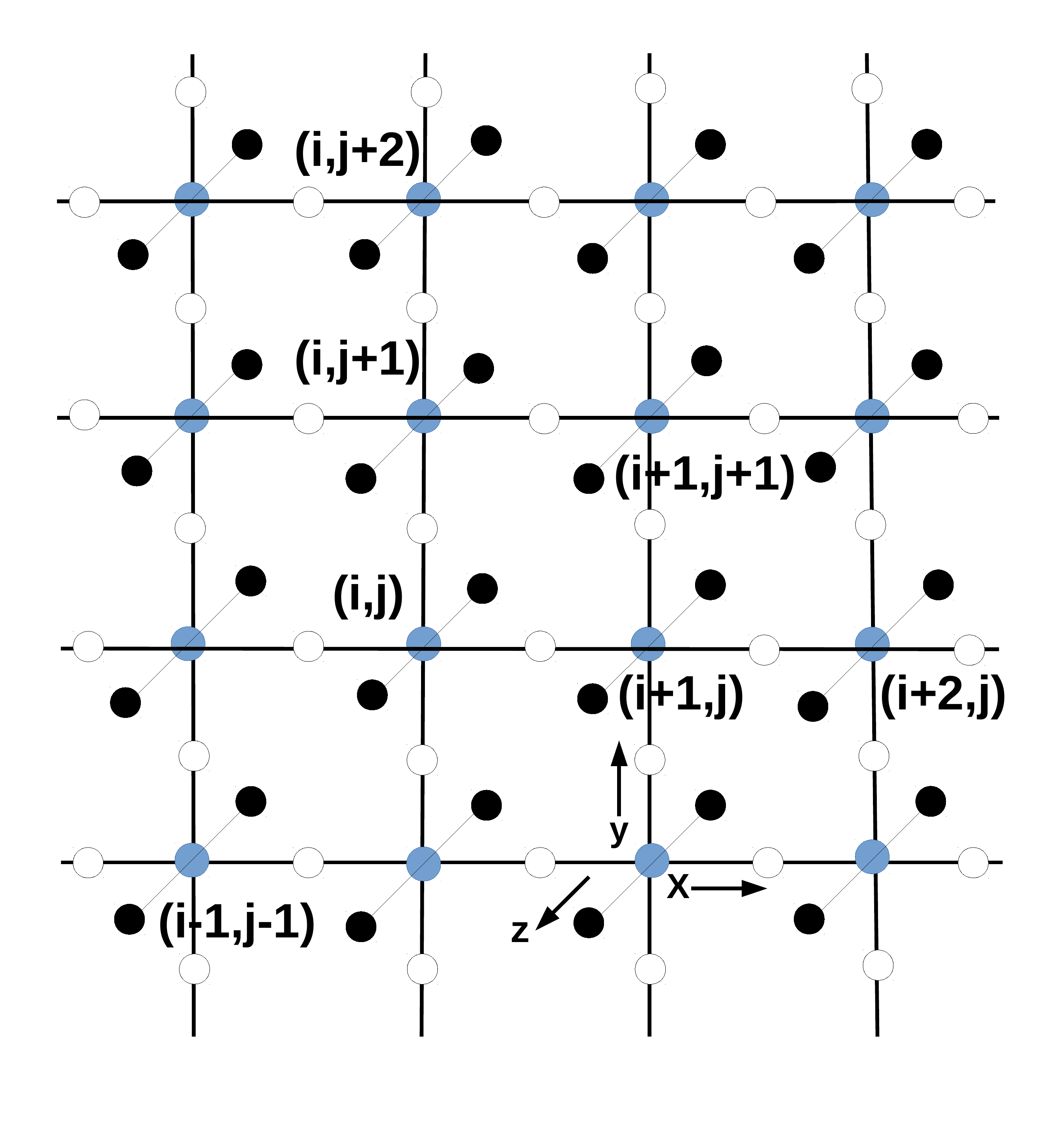}
\caption{(Color online) Schematic diagram for 
a 2D cooperative breathing mode (CBM) system. Hopping sites for holes
 are represented by blue solid
circles, in-plane oxygen atoms (participating
in the CBM) by black empty circles, and 
non-cooperative out-of-plane oxygen atoms
by solid black circles.}
\label{fig:2D_CBM}
\end{figure}

In this section we focus  on the analytical treatment 
of the effective Hamiltonian 
which will be used 
for numerical simulation.
We are working with a 2D version of the pervovskite manganite system  which has 
Mn-O-Mn bonds along the $x$ and $y$ directions. We
have  $\rm e_g$ electrons (or holes) interacting with the oxygen atoms.
 We have restricted our analysis to a system of fermions interacting with the
oxygens in the $xy$-plane via cooperative breathing mode
and with the out-of-plane  $z$-direction oxygens through 
non-cooperative breathing mode as depicted in Fig. \ref{fig:2D_CBM}.

Apart from the itinerant $e_{\rm g}$ electrons,
we also have a $ t_{\rm 2g}$ localized core-spin background;
 the large $S=2$ spin (comprising of contribution from a $e_{\rm g}$ spin and three  $ t_{\rm 2g}$ spins)
 at each site is considered classical.
Thus 
the Hamiltonian of such a system has four interactions:
the kinetic energy of the fermions, the fermion-lattice
coupling energy, the lattice energy, and the  spin-spin 
interaction energy.
\begin{eqnarray}
 H = H_{\rm KE} + H_{\rm int} + H_{\rm lat} + H_{\rm SE} .
\end{eqnarray}
Here, 
\begin{eqnarray}
 H_{\rm KE} = -t \sum\limits_{i,j} \bigg[\cos\left({{{\theta_{i,j;i+1,j}}\over2}}\right) {{d^{\dagger}}_{i+1,j} d_{i,j}}+\nonumber\\
 \cos\left({{{\theta_{i,j;i,j+1}}\over2}}\right) {{d^{\dagger}}_{i,j+1} d_{i,j}}+ {\rm H.c.}\bigg ] ,
\end{eqnarray}
and $d_{i,j} ({d^{\dagger}}_{i,j})$ represents the annihilation
 (creation) operator  for the
fermion at site $(i,j)$; $t$ is the hopping 
amplitude for the fermion;
the hopping process is modified by
 $\theta$, the angle between  two  spins at NN
sites.\cite{gennes,Izyumov}
The second term represents the interaction between the fermions and the
quantum phonons in the system and is expressed as \cite{cbm2d}
\begin{eqnarray}
 H_{\rm int}= &-&g\omega_{0} \sum\limits_{i,j} \Big \{ \big ({a^{\dagger}}_{x;i,j}+ a_{x;i,j} \big )(n_{i,j} - n_{i+1,j})\nonumber\\ 
           &+& \big ({b^{\dagger}}_{y;i,j}+ b_{y;i,j} \big )(n_{i,j} - n_{i,j+1})\nonumber\\
           &+&\gamma \big ({c^{\dagger}}_{z;i,j}+ c_{z;i,j}\big )n_{i,j} \Big \} ,
\end{eqnarray}
where $\gamma = \sqrt{2}$, 
$g$ is the electron-fermion coupling constant,
 $\omega_0$ is the optical-phonon frequency, and  $n_{i,j} = d^{\dagger}_{i,j} d_{i,j}$.
The displacement of the oxygen atom that is adjacent to the
site $(i,j)$  in the positive $x$ [$y$] direction
is given by ${({a^{\dagger}}_{x;i,j}+ a_{x;i,j})}\over{\sqrt{2m\omega_{0}}}$
\bigg [${({b^{\dagger}}_{y;i,j}+ b_{y;i,j})}\over{\sqrt{2m\omega_{0}}}$ \bigg ].
In the $z$ direction, 
the relative displacement of the two oxygen atoms next to site $(i,j)$ is denoted by
${({c^{\dagger}}_{z;i,j}+ c_{z;i,j})}\over{\sqrt{2m\omega_{0}/2}}$ with $m/2$ being the 
reduced mass of the oxygen pair.
Next, the lattice
energy due to 
quantum harmonic oscillators 
is given by
\begin{eqnarray}
\!\!\!\!\!\!\!\! H_{\rm lat} =  \sum\limits_{i,j} \Big ({a^{\dagger}}_{x;i,j} a_{x;i,j} +{b^{\dagger}}_{y;i,j} b_{y;i,j} +\eta {c^{\dagger}}_{z;i,j} c_{z;i,j} \Big ) ,
\end{eqnarray}
with $\eta$  being set to be 1.

Now, to arrive at an effective Hamiltonian which
can be expressed solely in terms of fermionic operators,
we take resort to an analytic approach similar 
to that described in \cite{cbm2d}. For large electron-phonon coupling
and restricting the system to the
non-adiabatic regime, $t/\omega_{0} \lesssim 1$, the above Hamiltonian 
$H$ is subject to a canonical transformation (i.e., modified
Lang-Firsov transformation) to produce an unperturbed 
part $H_0$ and the perturbation term 
$ H_1$. 
To obtain an effective Hamiltonian, we perform second-order
perturbation theory (as in Refs. \onlinecite{cbm2d,rpsy})
and obtain
\begin{eqnarray}
H_{\rm eff} &= &-E_p \sum\limits_{i,j} n_{i,j} + 2V_p \sum\limits_{i,j} \Big ({n_{i,j}n_{i+1,j}+n_{i,j}n_{i,j+1}} \Big )\nonumber\\
          &&+ te^{-(E_p + V_p)/\omega_{0}} \sum\limits_{i,j} \bigg[\cos\left({{{\theta_{i,j;i+1,j}}\over2}}\right){d^{\dagger}_{i+1,j} d_{i,j}}\nonumber\\ 
          &&+ \cos\left({{{\theta_{i,j;i,j+1}}\over2}}\right){d^{\dagger}_{i,j+1} d_{i,j}} + {\rm H.c.} \bigg ]\nonumber\\
          &&+H^{(2)} + H_{\rm SE} ,
\label{h_eff1}
\end{eqnarray}
where the polaronic energy $ E_p = \big (4+\gamma^{2} \big ) g^2 \omega_0 = 6g^2 \omega_0$ and 
the nearest-neighbor repulsion energy
$ V_p = g^2 \omega_0$; the second order perturbation theory yields the term $\rm H^{(2)}$.
The small parameter of the perturbation theory is $\sim\Big [{{t^2}\over{2(E_p + V_p)}\omega_0}\Big ]^{{1}\over{2}}$
as derived in Ref. \onlinecite{A_Dey}.
Now, the effective hopping term $te^{-(E_p + V_p)/\omega_{0}} \ll \omega_0$.
For large $g$, 
the effective hopping term will be very small compared to the other terms in $H_{\rm eff} - H_{\rm SE}$.
Hence, we ignore the kinetic energy of 
the system and treat the system as made up of carriers
that are localized due to disorder. Then, we are justified in treating the problem 
entirely classically with physics being governed by the
dominant potential energy terms in the
effective Hamiltonian.
The first term of $H_{\rm eff}$ can as well be 
represented in terms of electron-hole attraction instead of 
fermion-fermion repulsion. In general
\begin{eqnarray}
V_p \sum\limits_{i,j,\delta} n_{i,j} n_{i+\delta,j} &=& -V_p \sum\limits_{i,j,\delta} n_{i,j} ({1-n_{i+\delta,j}}) \nonumber\\ 
&&+V_p\sum\limits_{i,j,\delta} n_{i,j} .
\label{v_p}
\end{eqnarray}
This formalism adds a constant energy term $V_p\sum\limits_{i,j,\delta} n_{i,j}$ to the
Hamiltonian and thus does not change 
the physics of the problem. 
Then, using  Eq. (\ref{v_p}) we can rewrite Eq. (\ref{h_eff1}) as
\begin{eqnarray}
 H_{\rm eff} &=&  -E_p \sum\limits_{i,j} n_{i,j} - 2V_p \sum\limits_{i,j} \Big \{n_{i,j}(1-n_{i+1,j})\nonumber\\ 
 &&+ n_{i,j}(1-n_{i,j+1}) \Big \}+4V_p \sum\limits_{i,j} n_{i,j} \nonumber\\
 &&+ H^{(2)} + H_{\rm SE} .
 \label{h_eff2}
\end{eqnarray}
The 
convention  we will  
use throughout the paper is that 
$n_{i,j}$ will represent number 
density of a hole   at the lattice site $(i,j)$ of the system.
To calculate $ H^{(2)}$, we go through an  algebra similar to that mentioned in Appendix A of Ref. \onlinecite{cbm2d} and arrive at a 
nearest-neighbor repulsion term  corresponding to the process where a 
particle in 2D virtually hops to its NN and comes back.
When a hole at site $(i,j)$ hops 
to its NN site, such as  $(i+1,j)$, and comes back,
we need to keep track of the occupancy of 
the three relevant 
nearest-neighbor sites of the intermediate site $(i+1,j)$, i.e., the occupancy of the three sites $(i+2,j)$ , $(i+1,j+1)$ and $(i+1,j-1)$.
Depending on how many of these three sites 
are filled, the coefficient for 
the hopping-and-returning process 
will be modified.

Clearly, there are four such possibilities for the coefficients and they will be considered below.

\begin{widetext}
\subsection{Three NN sites of the intermediate site are filled by electrons.}

 In Fig. \ref{fig:NN_all_empty},  when the intermediate site containing an electron
 is surrounded by a hole and three electrons,  we depict the hole  
at site $(i,j)$ 
hopping to its NN site (the intermediate site) 
and returning back.
The intermediate site can be any of the four NNs
of the originating site $(i,j)$.
A schematic view of the four possibilities
is shown in Fig. \ref{fig:NN_all_empty}.

 When a hole is at $(i,j)$, its energy is equal to $- E_p$.
  The oxygen atoms on both the sides of the 
 initial site are attracted by the hole on the initial site 
 and hence are pulled towards the hole. When the hole
 virtually hops to the intermediate site, its energy is equal
 to $ E_p + 2V_p$ because the oxygen distortions remain unchanged;  in the energy of the intermediate state, $E_p$
arises due to the distortion without the hole whereas the 
 extra energy $2V_p$ (equal in magnitude to the
 NN repulsion energy between two holes) results
 due to displacing the oxygen atoms towards the
 initial site and away from the hole.
 Hence, change in the energy when the hole jumps 
 from the originating site to the intermediate site is
 equal to 
  $2E_p + 2V_p $. 
 Thus, the coefficient of the second order perturbation
 term turns out to be 
 ${{t^2}\over{2E_p + 2V_p}}$ and the contribution to $H^{(2)}$ from all the possibilities corresponding to Fig. \ref{fig:NN_all_empty} is given by
\begin{eqnarray}
H^{(2)}_1 &= &-{{t^2}\over{(2E_p + 2V_p)}}\sum\limits_{i,j} \Bigg[{\cos^2\left({{\theta_{i,j;i+1,j}}\over2}\right)}\Big\{n_{i,j}(1-n_{i+1,j})
 (1-n_{i+2,j}) (1-n_{i+1,j+1})(1-n_{i+1,j-1})\Big \} \nonumber \\
&& +{\cos^2\left({{\theta_{i,j;i-1,j}}\over2}\right)}\Big \{n_{i,j}(1-n_{i-1,j})(1-n_{i-2,j})
 (1-n_{i-1,j+1})(1-n_{i-1,j-1})\Big \}\nonumber\\
&&+{\cos^2\left({{\theta_{i,j;i,j+1}}\over2}\right)}\Big \{n_{i,j}(1-n_{i,j+1})(1-n_{i,j+2})
 (1-n_{i-1,j+1})(1-n_{i+1,j+1})\Big\}\nonumber\\
&&+{\cos^2\left({{\theta_{i,j;i,j-1}}\over2}\right)}\Big\{n_{i,j}(1-n_{i,j-1})(1-n_{i,j-2})
 (1-n_{i-1,j-1})(1-n_{i+1,j-1})\Big\}\Bigg] .
 \label{H_2_1}
\end{eqnarray}

\begin{figure}[t]
 \includegraphics[width=7.0cm]{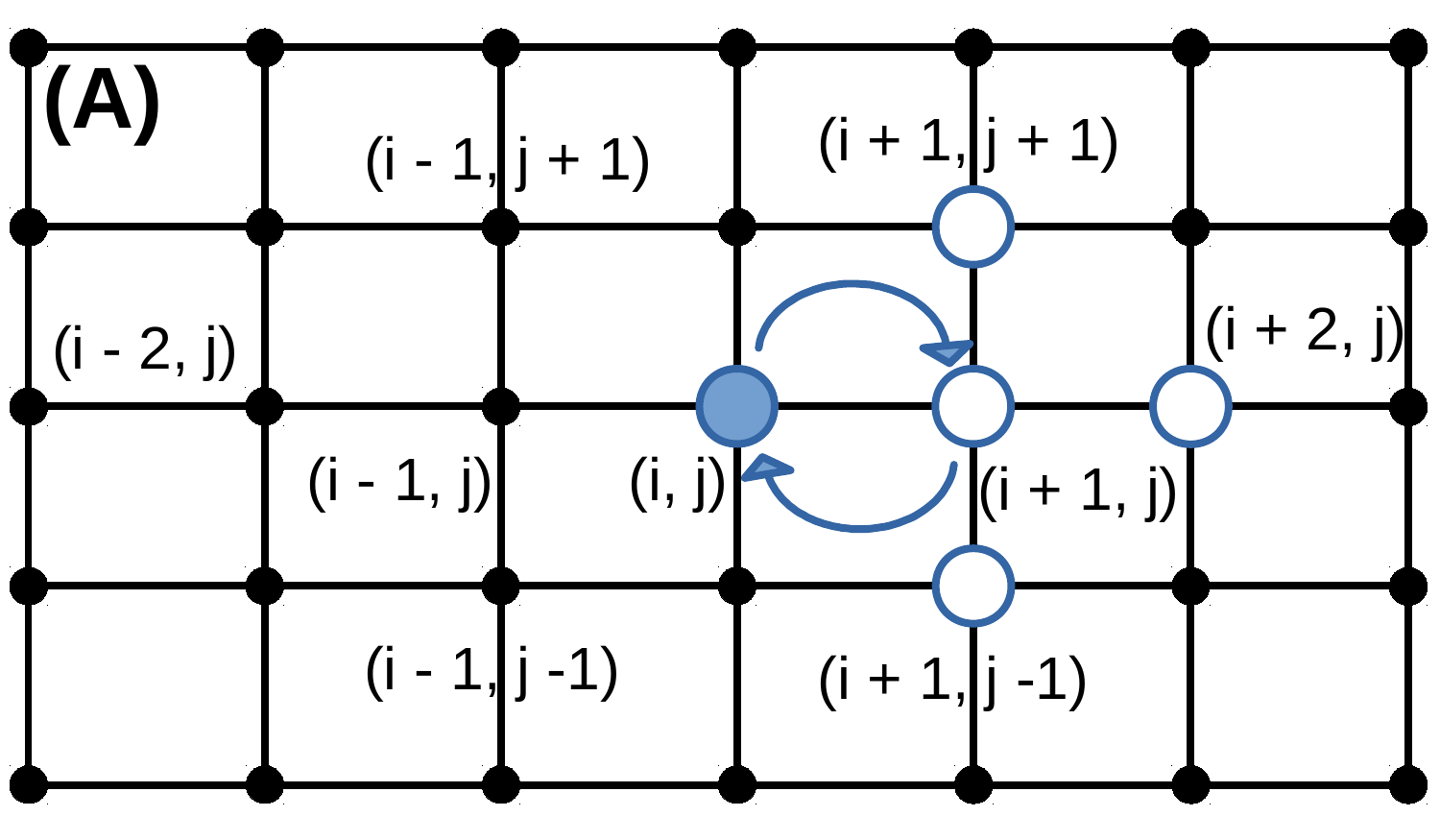}~~
 \includegraphics[width=7.0cm]{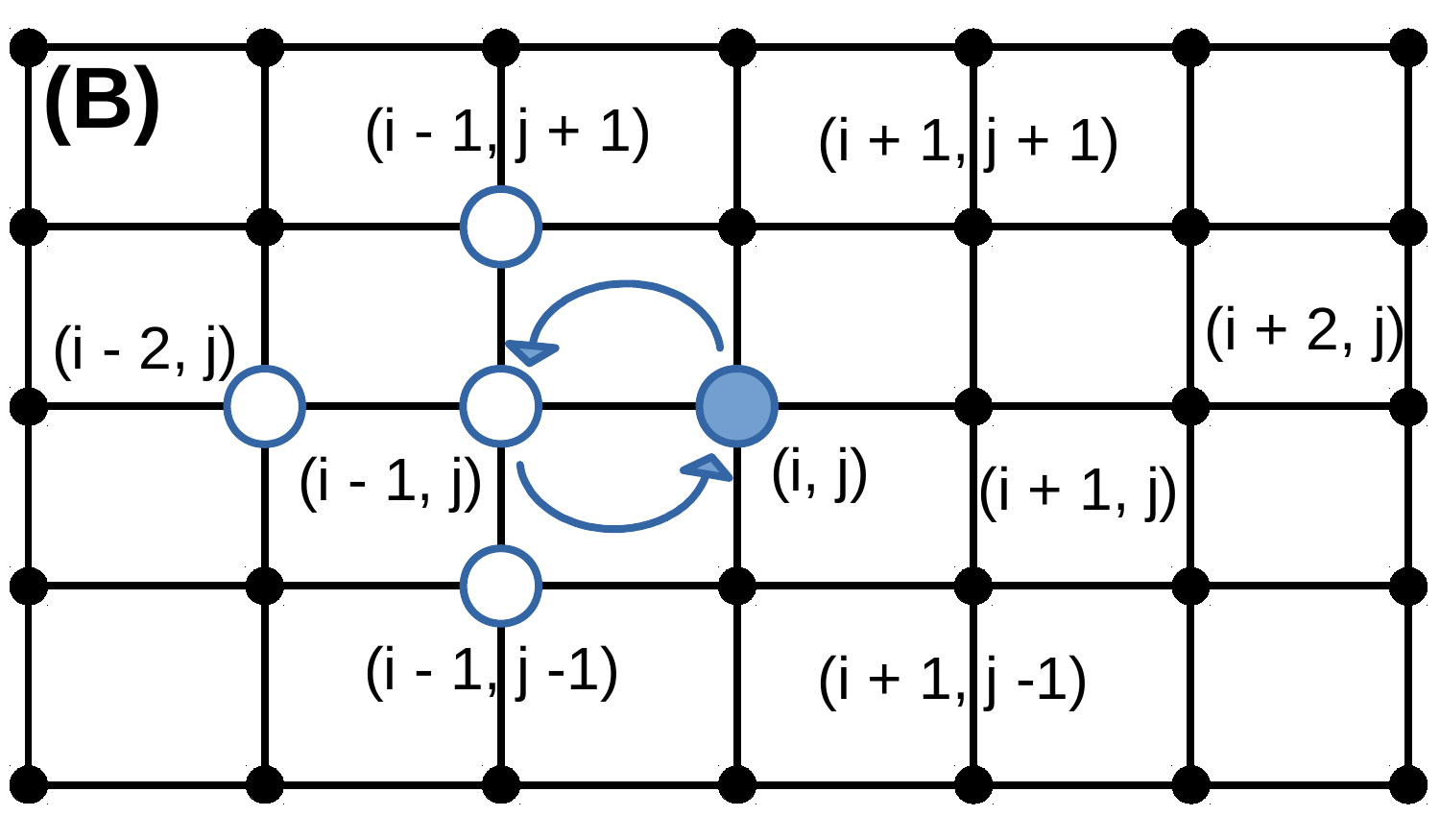}\\
 \includegraphics[width=7.0cm]{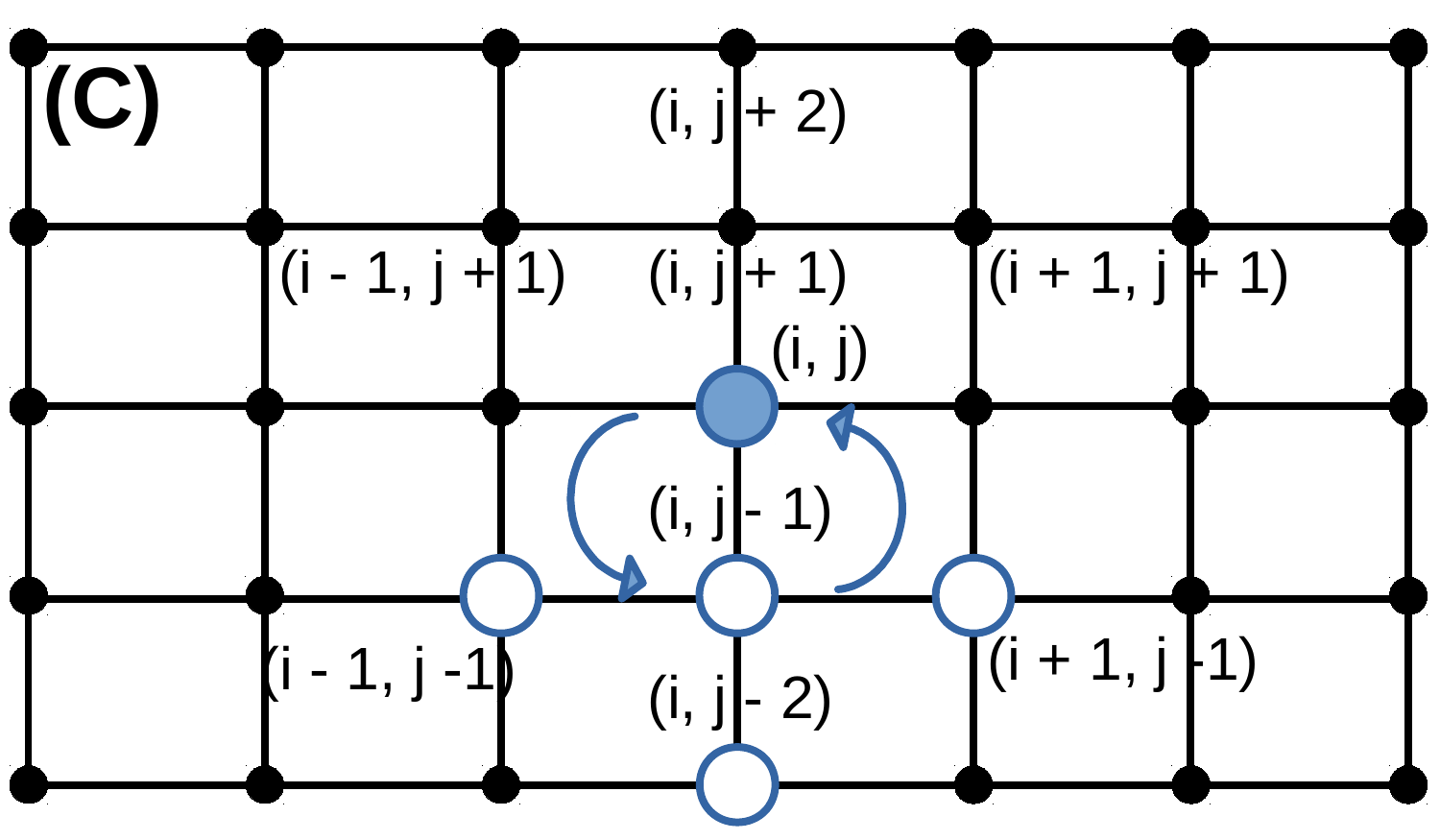}~~
 \includegraphics[width=7.0cm]{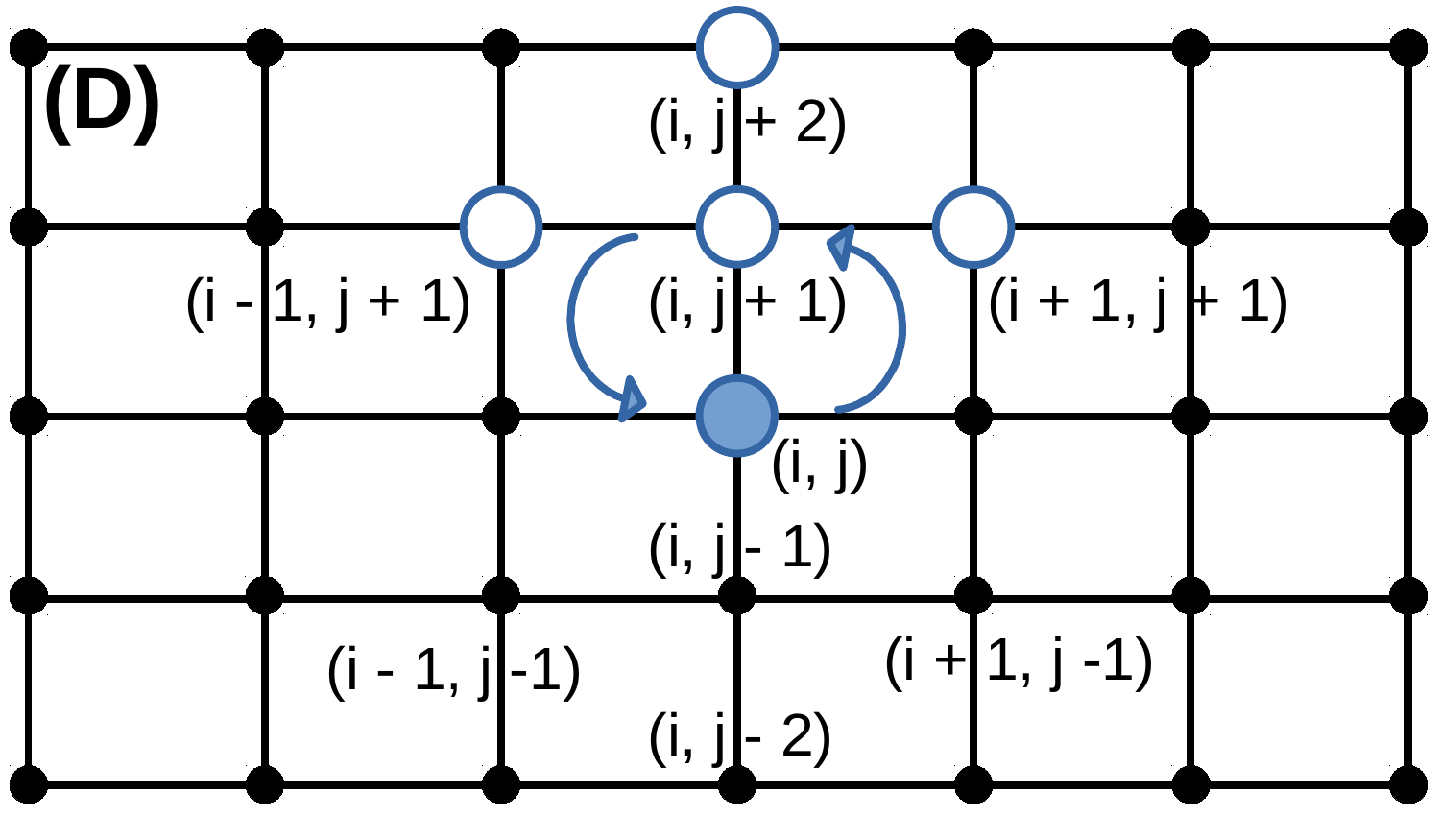}
\caption{(Color online) Schematic diagram for the four possibilities of a hole, at 
an originating site $(i,j)$,
hopping to its NN site (the intermediate site) 
and coming back
(when three NN sites of the intermediate site are
occupied by electrons):
(A) hole at $(i,j)$ hops to
its right NN at $(i+1,j)$ and comes back; (B) hole at $(i,j)$ jumps to
its left NN at $(i-1,j)$ and returns back; (C) hole at $(i,j)$ jumps to
its downward NN at $(i,j-1)$ and comes back; (D) hole at $(i,j)$ hops to
its upper NN at $(i,j+1)$ and returns. A hole is represented by a 
blue solid circle and a particle (i.e., electron) by a blue empty circle.
All 
lattice
sites that  are not relevant to the consideration are represented by black solid circles.}
 \label{fig:NN_all_empty}
\end{figure}

\subsection{Any two of NN sites of the intermediate site is filled.}

In Fig. \ref{fig:NN_two_filled}, we depict the three possibilities corresponding to  
a hole at a site $(i,j)$
hopping to its NN site (the intermediate site) 
and returning; here, any two of the NN sites of the 
intermediate site are occupied by electrons.
 Henceforth,
we will show all the counterpart processes
 of Fig. \ref{fig:NN_all_empty}(A) (considering
these as representative diagrams) for various possibilities.
Similar processes, which will not be shown here,
also occur for
 Fig. \ref{fig:NN_all_empty}(B), Fig. \ref{fig:NN_all_empty}(C), and Fig. \ref{fig:NN_all_empty}(D).

 When the hole
 virtually hops to the intermediate site, its energy is equal
 to $ E_p + 4V_p$; here, an extra repulsion of $2V_p$ is generated
 due to the occupancy of any one of the NN site 
 of the intermediate site by a hole. 
 Then, the coefficient of the second order perturbation
 term is 
 ${{t^2}\over{2E_p + 4V_p}}$ and the contribution to $H^{(2)}$ from all the possibilities, similar to and corresponding to  
 Fig. \ref{fig:NN_two_filled}, is given by

\begin{eqnarray}
 H^{(2)}_2 &= & -{{t^2}\over{(2E_p + 4V_p)}}\sum\limits_{i,j} \Bigg [{\cos^2\left({{\theta_{i,j;i+1,j}}\over2}\right)}\Big\{ n_{i,j}
(1-n_{i+1,j})n_{i+2,j} (1-n_{i+1,j+1})(1-n_{i+1,j-1})\nonumber\\
&&+n_{i,j}(1-n_{i+1,j})(1-n_{i+2,j})n_{i+1,j+1}(1-n_{i+1,j-1})
+n_{i,j}(1-n_{i+1,j})(1-n_{i+2,j})(1-n_{i+1,j+1})n_{i+1,j-1}\Big\} \nonumber\\
&&+{\cos^2\left({{\theta_{i,j;i-1,j}}\over2}\right)}\Big \{ n_{i,j}(1-n_{i-1,j})n_{i-2,j}
(1-n_{i-1,j+1})(1-n_{i-1,j-1})\nonumber\\
&&+n_{i,j}(1-n_{i-1,j})(1-n_{i-2,j})n_{i-1,j+1}(1-n_{i-1,j-1})
+n_{i,j}(1-n_{i-1,j})(1-n_{i-2,j})(1-n_{i-1,j+1})n_{i-1,j-1}\Big \} \nonumber\\
&&+{\cos^2\left({{\theta_{i,j;i,j+1}}\over2}\right)}\Big\{ n_{i,j}(1-n_{i,j+1})n_{i,j+2}
(1-n_{i-1,j+1})(1-n_{i+1,j+1})\nonumber\\
&&+n_{i,j}(1-n_{i,j+1})(1-n_{i,j+2})n_{i-1,j+1}(1-n_{i+1,j+1})
+n_{i,j}(1-n_{i,j+1})(1-n_{i,j+2})(1-n_{i-1,j+1})n_{i+1,j+1}\Big \}\nonumber\\
&&+{\cos^2\left({{\theta_{i,j;i,j-1}}\over2}\right)}\Big \{ n_{i,j}(1-n_{i,j-1})n_{i,j-2}
(1-n_{i-1,j-1})(1-n_{i+1,j-1})\nonumber\\
&&+n_{i,j}(1-n_{i,j-1})(1-n_{i,j-2})n_{i-1,j-1}(1-n_{i+1,j-1})\nonumber\\
&&+n_{i,j}(1-n_{i,j-1})(1-n_{i,j-2})(1-n_{i-1,j-1})n_{i+1,j-1}\Big\} \Bigg] .
\label{H_2_2}
\end{eqnarray}

\begin{figure}
 \includegraphics[width=7.0cm]{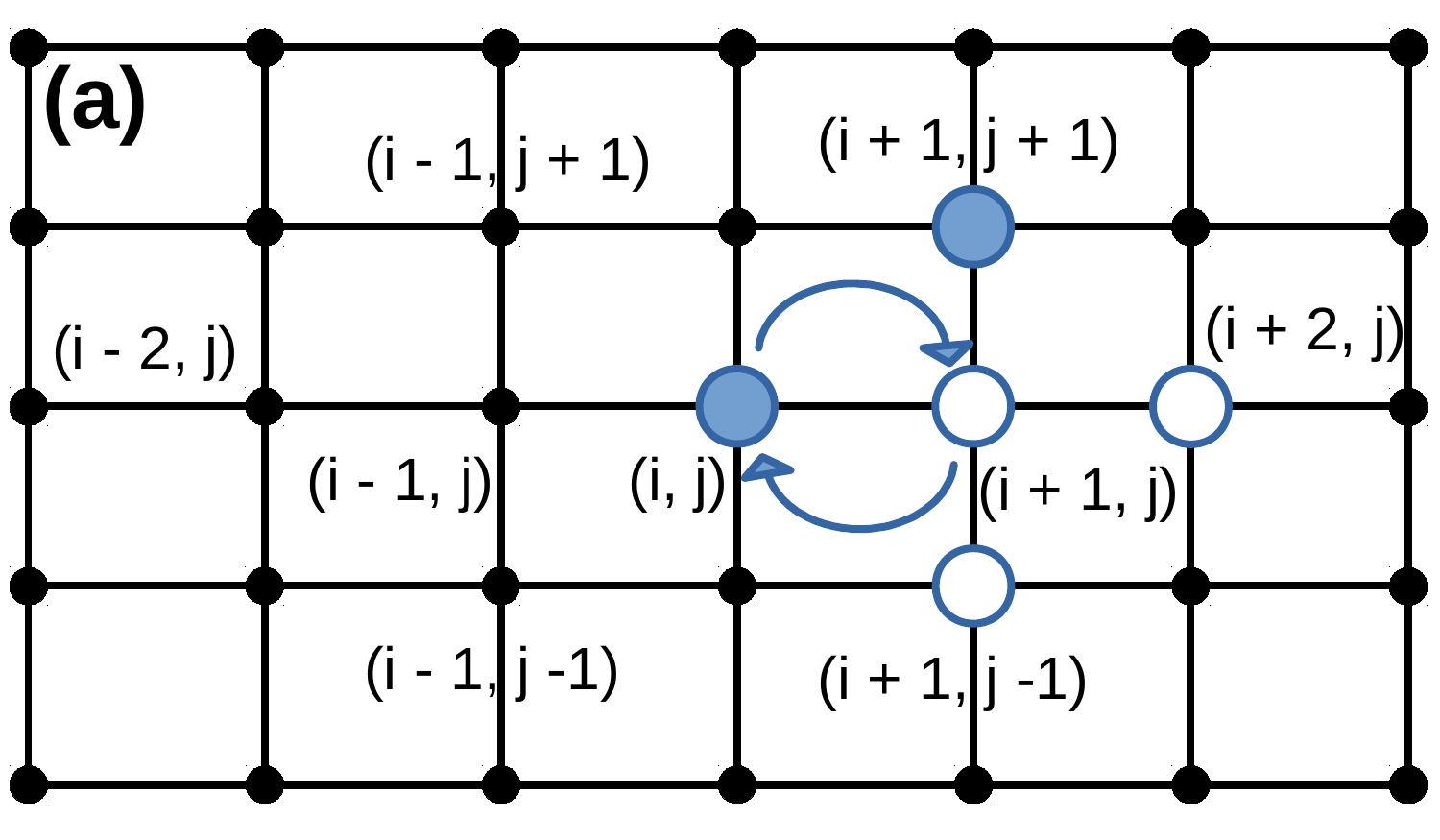}~
 \includegraphics[width=7.0cm]{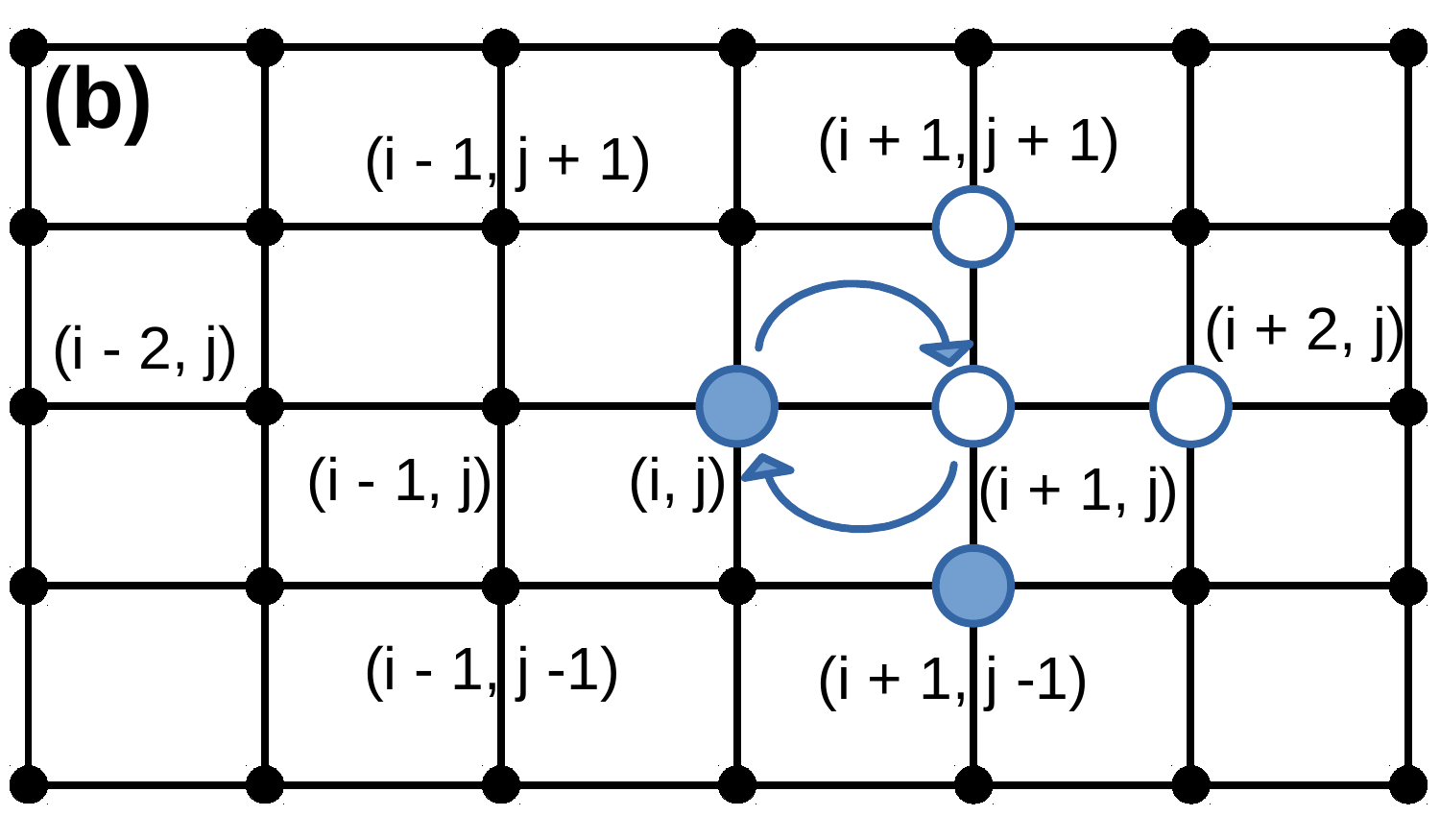} \\
 \includegraphics[width=7.0cm]{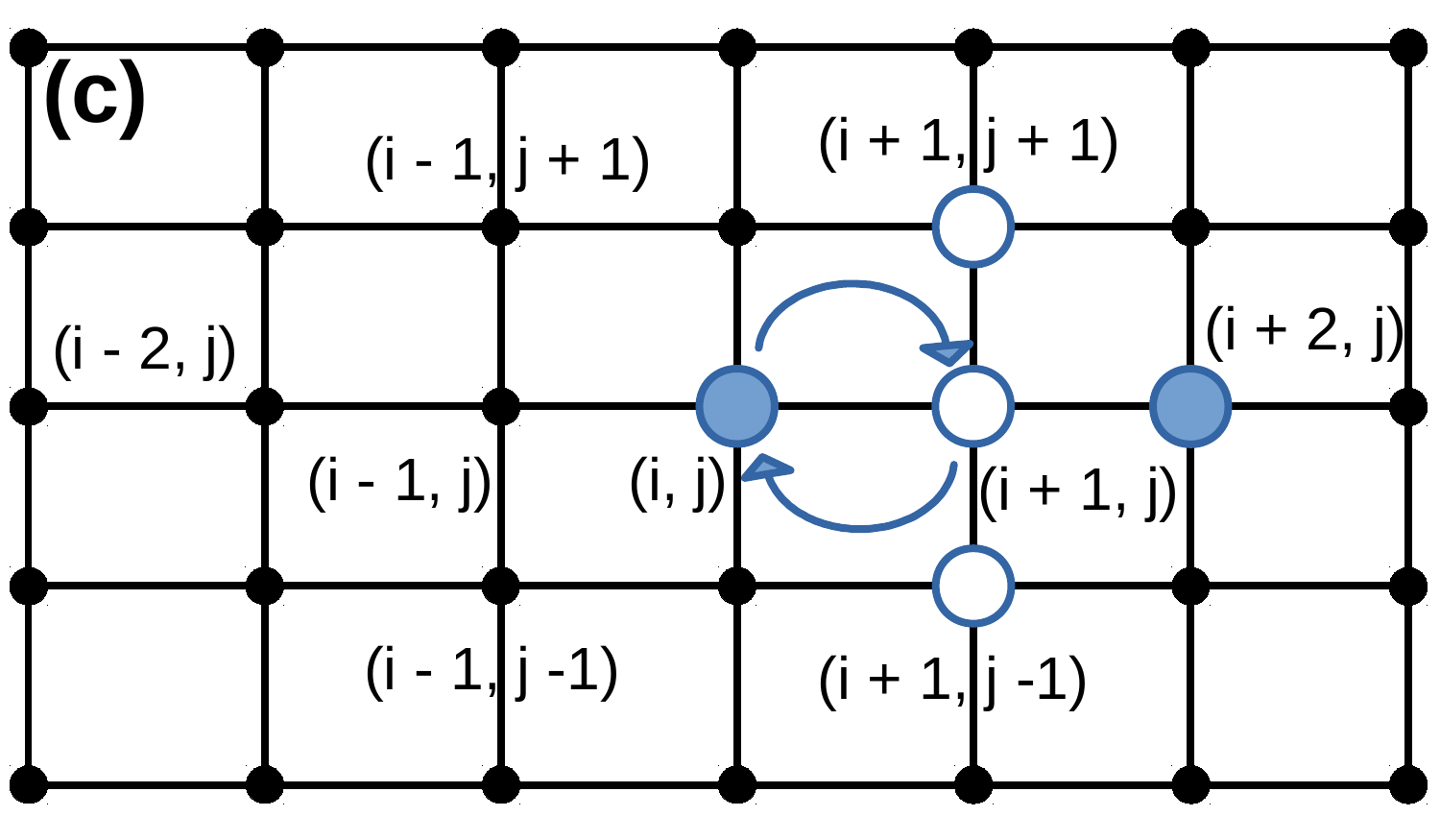}
 \caption{(Color online) Schematic diagram for a hole at 
an originating site $(i,j)$
hopping to its NN site (the intermediate site) 
and returning back (when any two of the NN sites of the 
intermediate site are occupied by electrons). Representation of
a hole at $(i,j)$ jumping to
its right NN at $(i+1,j)$ and coming back
when  holes occupy (a) right and downward NNs of the intermediate 
site; (b) right and upward NNs of the intermediate 
site; (c) upward and downward NNs of the intermediate 
site.
A hole is depicted by a 
blue solid circle and a particle by a blue empty circle.
All  lattice
sites that are not relevant to the consideration are represented by black solid circles.}
\label{fig:NN_two_filled}
\end{figure}

\subsection{Any one of NN sites of the intermediate site has an electron.}

In Fig. \ref{fig:NN_one_filled}, three possibilities
have been shown for the process where a hole jumps to
an intermediate site and comes back; here, any one of the 
NNs of the intermediate site is filled by an electron.
Extending the logic given above to the present case, the coefficient of the second order perturbation
 term is 
 ${{t^2}\over{2E_p + 6V_p}}$ and the contribution to $H^{(2)}$ from all the possibilities, similar to and corresponding to  
 Fig. \ref{fig:NN_one_filled}, is given by
\begin{eqnarray}
 H^{(2)}_3 &= &-{{t^2}\over{(2E_p + 6V_p)}}\sum\limits_{i,j} \Bigg [{\cos^2\left({{\theta_{i,j;i+1,j}}\over2}\right)}\Big\{ n_{i,j}(1-n_{i+1,j})
n_{i+2,j} n_{i+1,j+1}(1-n_{i+1,j-1})\nonumber\\
&&+n_{i,j}(1-n_{i+1,j})(1-n_{i+2,j})n_{i+1,j+1}n_{i+1,j-1}
+n_{i,j}(1-n_{i+1,j})n_{i+2,j}(1-n_{i+1,j+1})n_{i+1,j-1}\Big\} \nonumber\\
&&+{\cos^2 \left({{\theta_{i,j;i-1,j}}\over2}\right)}\Big\{ n_{i,j}(1-n_{i-1,j})n_{i-2,j}n_{i-1,j+1}(1-n_{i-1,j-1})\nonumber\\
&&+n_{i,j}(1-n_{i-1,j})(1-n_{i-2,j})n_{i-1,j+1}n_{i-1,j-1}
+n_{i,j}(1-n_{i-1,j})n_{i-2,j}(1-n_{i-1,j+1})n_{i-1,j-1}\Big \} \nonumber\\
&&+{\cos^2 \left({{\theta_{i,j;i,j+1}}\over2}\right)}\Big \{ n_{i,j}(1-n_{i,j+1})n_{i,j+2}(1-n_{i-1,j+1})n_{i+1,j+1}\nonumber\\
&&+n_{i,j}(1-n_{i,j+1})(1-n_{i,j+2})n_{i-1,j+1}n_{i+1,j+1}
+n_{i,j}(1-n_{i,j+1})n_{i,j+2}n_{i-1,j+1}(1-n_{i+1,j+1})\Big\} \nonumber\\
&&+{\cos^2 \left({{\theta_{i,j;i,j-1}}\over2}\right)}\Big \{ n_{i,j}(1-n_{i,j-1})n_{i,j-2}n_{i-1,j-1}(1-n_{i+1,j-1})\nonumber\\
&&+n_{i,j}(1-n_{i,j-1})(1-n_{i,j-2})n_{i-1,j-1}n_{i+1,j-1}
+n_{i,j}(1-n_{i,j-1})n_{i,j-2}(1-n_{i-1,j-1})n_{i+1,j-1}\Big\} \Bigg]
\label{H_2_3}
\end{eqnarray}
\begin{figure}
 \includegraphics[width=7.0cm]{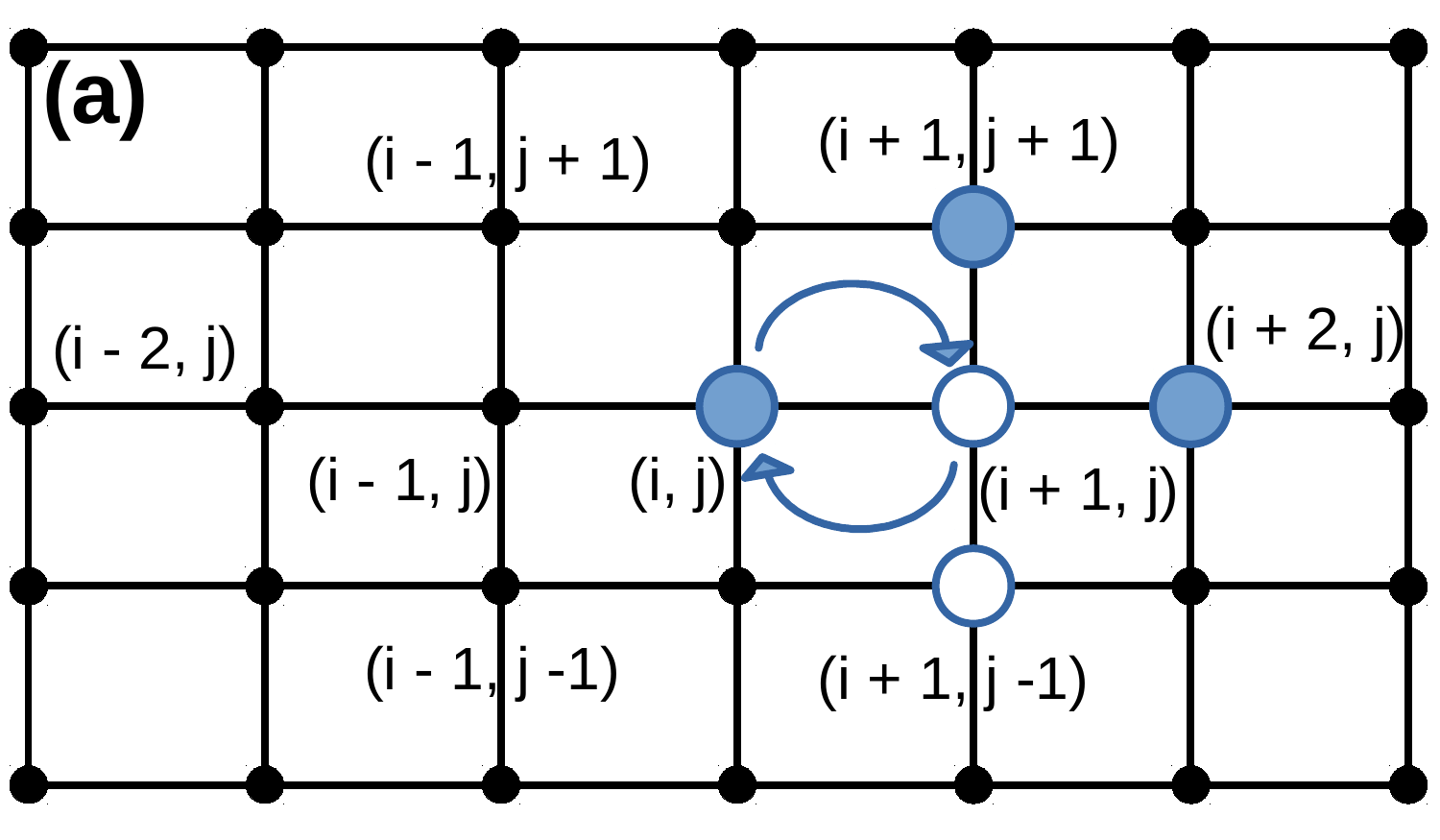}~~
 \includegraphics[width=7.0cm]{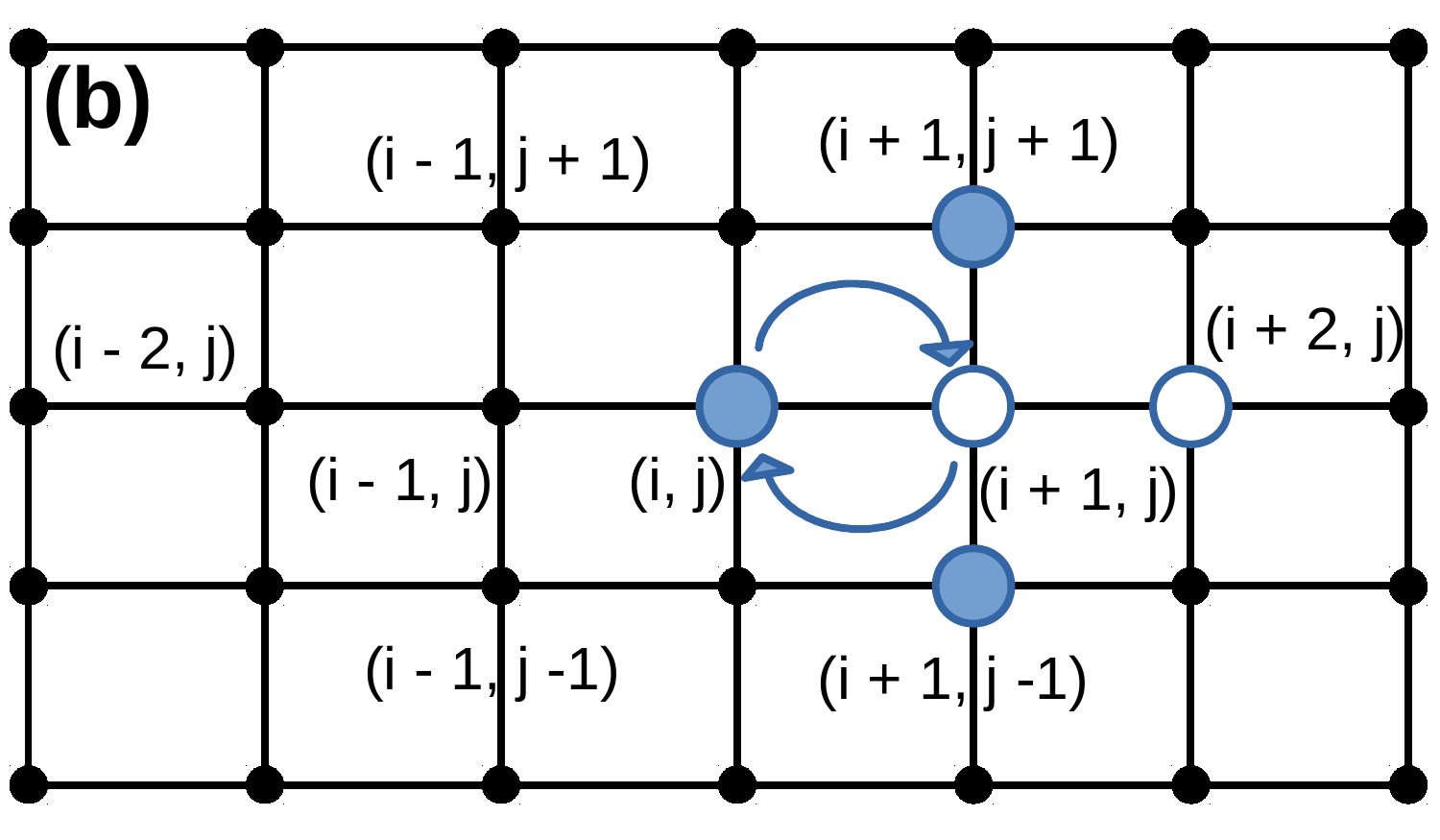}\\
 \includegraphics[width=7.0cm]{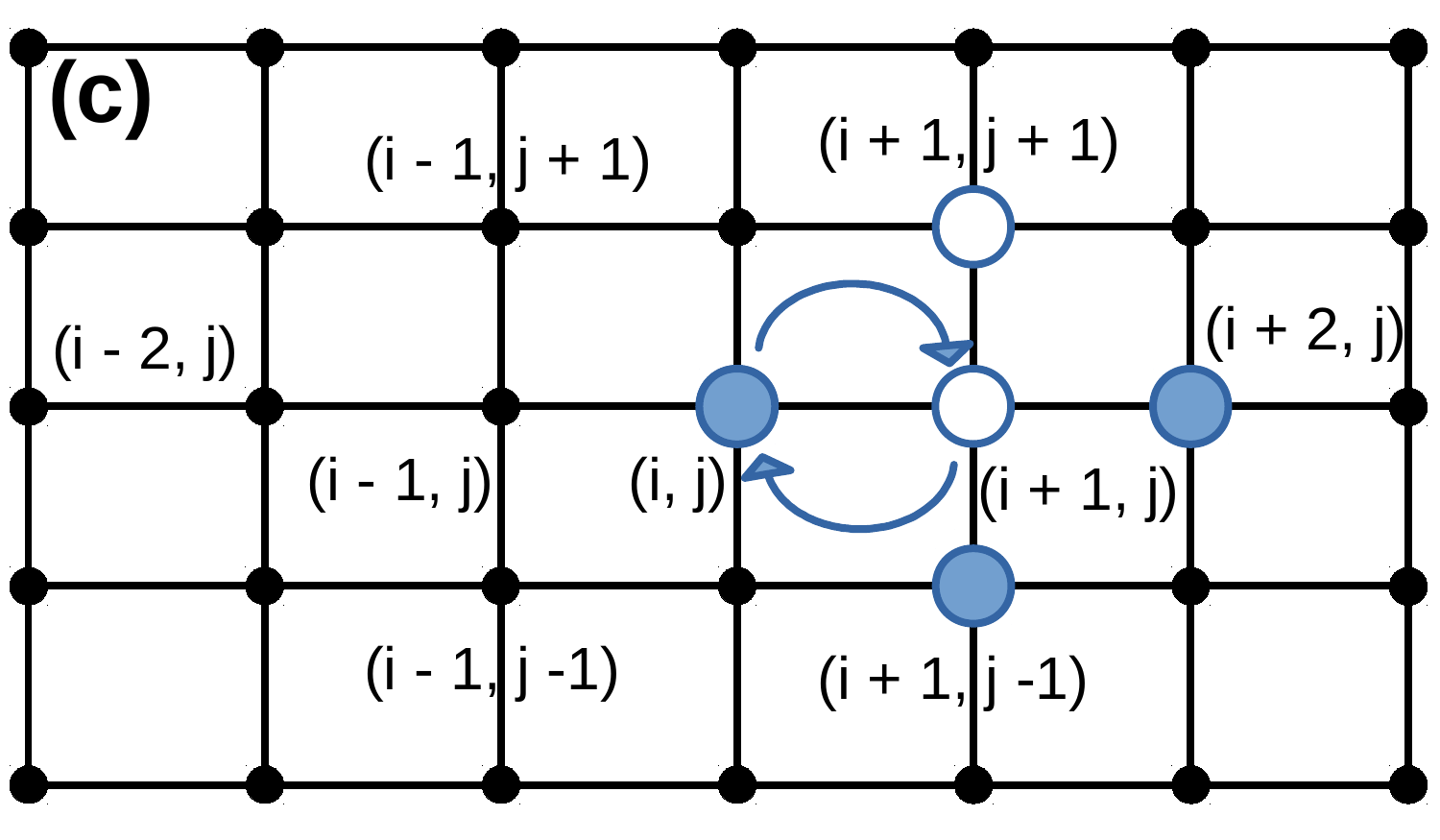}
 \caption{(Color online) Schematic diagram for a hole at 
an originating site $(i,j)$
hopping to its NN site (the intermediate site) 
and coming back (when any one of the NN sites of the 
intermediate site is occupied by an electron). Depiction of
a hole at $(i,j)$ jumping to
its right NN at $(i+1,j)$ and coming back
when a particle occupies (a) downward NN of the intermediate 
site; (b) right NN of the intermediate 
site; (c) upward NN of the intermediate 
site. A hole is represented by a 
blue solid circle whereas a particle by a blue empty circle.
All  lattice
sites that are not relevant to the consideration are indicated by black solid circles.}
 \label{fig:NN_one_filled}
\end{figure}

\subsection{All the 
NN sites of the intermediate site have holes}.

Here, for the situation where all the NN sites of the intermediate site have holes,
we depict in Fig. \ref{fig:NN_all_filled} a hole hopping
to an intermediate site and coming back.
Here, the coefficient of the second order perturbation
 term is 
 ${{t^2}\over{2E_p + 8V_p}}$ and the contribution to $H^{(2)}$ from all the possibilities, similar to and corresponding to  
 Fig. \ref{fig:NN_all_filled}, is given by
\begin{eqnarray}
 H^{(2)}_4 = -{{t^2}\over{(2E_p + 8V_p)}}\sum\limits_{i,j}\Bigg[{\cos^2 \left({{\theta_{i,j;i+1,j}}\over2}\right)}\Big\{n_{i,j}(1-n_{i+1,j})
 n_{i+2,j} n_{i+1,j+1}n_{i+1,j-1}\Big\}\nonumber\\
+{\cos^2 \left({{\theta_{i,j;i-1,j}}\over2}\right)}\Big\{n_{i,j}(1-n_{i-1,j})n_{i-2,j}
n_{i-1,j+1}n_{i-1,j-1}\Big\}\nonumber\\
+{\cos^2\left({{\theta_{i,j;i,j+1}}\over2}\right)}\Big\{n_{i,j}(1-n_{i,j+1})n_{i,j+2}
n_{i-1,j+1}n_{i+1,j+1}\Big\}\nonumber\\
+{\cos^2 \left({{\theta_{i,j;i,j-1}}\over2}\right)}\Big\{n_{i,j}(1-n_{i,j-1})n_{i,j-2}
n_{i-1,j-1}n_{i+1,j-1}\Big\} \Bigg] .
\label{H_2_4}
\end{eqnarray}

\begin{figure}[t]
\includegraphics[width=7.0cm]{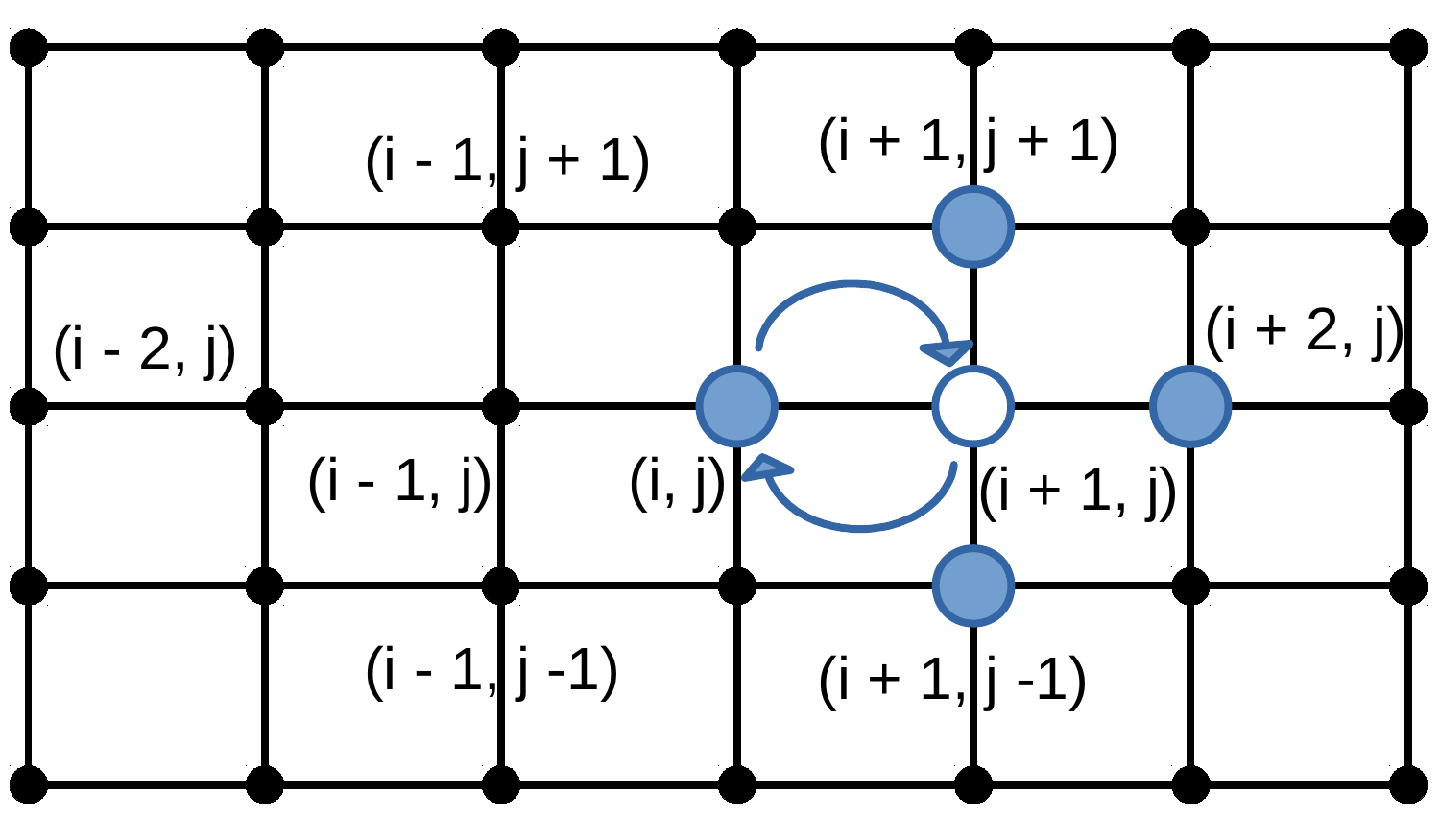}
\caption{(Color online) Schematic diagram for a hole at 
an originating site $(i,j)$
hopping to its NN site (the intermediate site) 
and coming back (when all the other three NN sites of the 
intermediate site are  occupied by electrons). Representation of
a hole at $(i,j)$ jumping to
its right NN at $(i+1,j)$ and coming back.
A hole is represented by a 
blue solid circle and a particle by a blue empty circle.
All  lattice
sites irrelevant to the analysis are represented by black solid circles.} 
\label{fig:NN_all_filled}
\end{figure} 
\end{widetext}

From the contributions $H^{(2)}_1$, $H^{(2)}_2$,
 $H^{(2)}_3$, and $\rm H^{(2)}_4$ obtained above,
 we express $H^{(2)}$ as
 \begin{eqnarray}
 H^{(2)} = H^{(2)}_1 + H^{(2)}_2 + H^{(2)}_3 + H^{(2)}_4 .
 \label{total H_2}
 \end{eqnarray}

Lastly, the superexchange \cite{anderson} term $\rm H_{SE}$ generates
A-AFM spin-spin exchange in manganites such as $\rm LaMnO_3$ and is given by
\begin{eqnarray}
 H_{\rm SE} = - J_{\rm x y} \sum\limits_{\langle i,j \rangle_{\rm x y}} \cos\left (\theta_{ij}\right )
 + J_{\rm z} \sum\limits_{\langle i,j \rangle_{\rm z}} \cos\left (\theta_{ij}\right ) .
 \label{H_se1}
\end{eqnarray}
It is important to note that, while the range of charge-charge interaction
is as far as NNNN, the range of spin-spin interaction is only NN.

\section{Calculation procedure}
\label{calc_proc}

For a numerical study, we consider a 2D lattice 
with periodic boundary conditions
in both directions.
We treat the problem fully classically
using the effective Hamiltonian, comprising of
the effective electron-phonon interaction 
(the charge-spin-coupled term) and the 
superexchange interaction (the  spin-spin interaction term),
as given by Eq. (\ref{h_eff2}).
We use classical Monte Carlo technique 
and make use of the standard 
Metropolis algorithm to update  
the charge configuration as well as the spin configuration
of the system. We follow a two-step 
procedure to arrive at the final charge and spin configurations.

Firstly, 
 to deal with problem of  charge configurations that correspond to local minima which are close in energy, we take 
resort to simulated annealing for the
charge degrees of freedom only. The spin variables 
are kept frozen since the energy scale for the charge interactions  is
much higher than the energy scale for superexchange interactions. 
Since we are working with 
low hole densities (i.e., between 0.1 and 0.3),
a large number of degenerate states will appear
in the charge spectrum. In order to obtain maximum number of
such degenerate configurations, we employ a three-step procedure at 
each temperature of the simulated 
annealing process to obtain the  optimized 
charge configurations. 
The primary step is a \textquotedblleft single-particle-exchange\textquotedblright~
 process where we choose any two sites at a
time---one sequentially and the other 
randomly---and exchange their number density values
provided they differ by 1. Physically we exchange a particle at 
a site with a hole at any other site.
The secondary step is a \textquotedblleft general-two-particle-exchange\textquotedblright~ 
process where any two random sites are 
selected with both being occupied by particles
and then their occupants  are
exchanged with another pair of randomly 
chosen sites both containing holes.
Thus we 
actually exchange two particles with two holes at a time. 
The final step---a \textquotedblleft plaquette-exchange\textquotedblright process---is a special case of 
the \textquotedblleft general-two-particle-exchange\textquotedblright 
mechanism. Here plaquette are chosen 
sequentially; if the difference in 
number densities between the two
diagonal pairs is 2, then the number densities
of the diagonals are exchanged.
At a particular temperature, 
to arrive at the final lowest energy 
charge configuration
at that temperature using 
Monte Carlo technique,
an initial random 
charge configuration 
(with a fixed number of 
particles) first goes through
$4\times 10^5$ steps of \textquotedblleft single-particle-exchange\textquotedblright;
then an equal number of steps involving  
\textquotedblleft general-two-particle-exchange\textquotedblright;
followed by 30 times the system size number 
\textquotedblleft plaquette-exchange\textquotedblright steps.

Secondly, using the charge profile  
generated by the three-step process,
we now  optimize  spin 
variables by taking an initial
random spin configuration and updating through 
the Metropolis algorithm.
The 
spins being large 
in magnitude, with $S=2$, are essentially 
classical spins with $\vec{S_i} = (\sin\theta_i \cos\phi_i, \sin\theta_i \sin\phi_i, \cos\theta_i)$.
While updating the spins, 
we consider the full Hamiltonian $ H_{\rm eff}$ and consider both
the charge and spin interaction energies. 
The  $\rm \cos(\theta)$ and $\rm \phi$ 
values are binned to fix the
orientation of the classical 
spin vector.
 We have allowed equally spaced
40 values of $\rm \cos(\rm \theta)$ in the interval $(-1,1)$ 
and 80 values of $\phi$ 
in the usual range of $(0,2\pi)$, thus totaling
to 
3200 different possibilities.
A sweep involves visiting all the lattice sites sequentially and
updating the spin orientation at each lattice site by the
 Metropolis  algorithm.
 The equilibrium number of sweeps required for medium 
(higher) temperatures is 
around $15\times 10^5$ ($6\times 10^5$), while another $15\times 10^5$
($6\times 10^5$)
sweeps are required for 
the thermal averaging of the 
total magnetization of the system. 
It is to be noted that for low
hole concentrations,  we have 
many degenerate
states. We calculate the magnetization for typically 10
degenerate configurations. 
The degenerate
states are chosen based on the 
charge optimization 
process only, fed to the full Hamiltonian $H_{\rm eff}$ containing
both charge and spin variables, 
and then energy is optimized
to obtain the total magnetization 
of all 
such states. The  magnetization/site 
of the system, that has been
plotted, is the magnetization/site
averaged over all the degenerate states for a 
particular filling of holes when spins normalized to unity.

We study the system for the bare hopping 
parameter values $t=0.2$ eV and $t=0.3$ eV.
Our calculations  take  the polaronic energy to be  $E_p = 0.43$ eV
and the nearest-neighbor repulsion energy to be $V_p = 0.07$ eV. 
Thus, we are in the regime of strong electron-phonon coupling characterized by
$(E_p+V_p)/\omega_0 >> 1$ with the optical phonon frequency $\omega_0$ value being given as 0.05 eV $< \omega_0 <$ 0.1 eV.
The superexchange energy 
coefficient $J_{\rm z} = 4.8$ meV \cite{khali,super_ex};
thus the superexchange energy is much smaller than the
electron-hole pair ferromagnetic interaction coupling
[$t^2/(2E_p+2V_p)$].
Furthermore, the ferromagnetic coupling $J_{\rm x y}$ = $1.4\times J_{\rm z}$.
Thus, the 
charge configuration can be assumed to remain
constant as the spins are optimized. 
The total magnetization of the system
is computed at various temperatures, with the highest temperature being 
$T =0.1t/k_B$  (i.e., about 330K for $t=0.3$ eV. Henceforth, $k_B$ will be set to unity for convenience..
The lowest temperature on the
other hand is $T =0.001t$ (i.e., about 3K for $t=0.3$ eV)
which is much smaller than $J_z$; 
thus, the system can 
be assumed to be in its ground 
state at $T =0.001t$.
{
Here, we should comment that above $T=0.03t$, the excited-state charge configurations also
begin to contribute to the magnetization}.

\section{Results and discussion}
We consider a 2D lattice 
of dimensions $6\times12$,
(i.e., with a total  of 72 sites) with
periodic boundary conditions in both 
directions; the number of rows being
$\rm l_x = 6$ and the number of columns 
being $\rm l_y = 12$. Each site represents 
an Mn ion consisting of an electron 
and a positive charge center.
We study the interplay between 
the electron-phonon interaction and 
the magnetic interaction of 
the  spins
[see  Eq. (\ref{h_eff2})]. 
As stated earlier 
(in Sec. \ref{intro} ), due to the 
the smallness of the kinetic energy in comparison to the
potential energy, 
the problem is treated fully classically.
Thus the holes are site-localized  and the system can be represented by a single state in the
occupation number basis with the number density at each site being
 either 1 or 0.
  Hence,
for strong electron-phonon interaction,
we have a fully insulating system
resembling a charge solid as shown in Fig. \ref{fig:Charge_configs}.
Using effective Hamiltonian in Eq. (\ref{h_eff2}),
we can simulate different observables in the system.
We study the variation of the total magnetization
of the system as a function of
hole doping in the pure manganite sample.

\subsection{A-AFM background}
\begin{figure}[]
 \includegraphics[width=8.0cm]{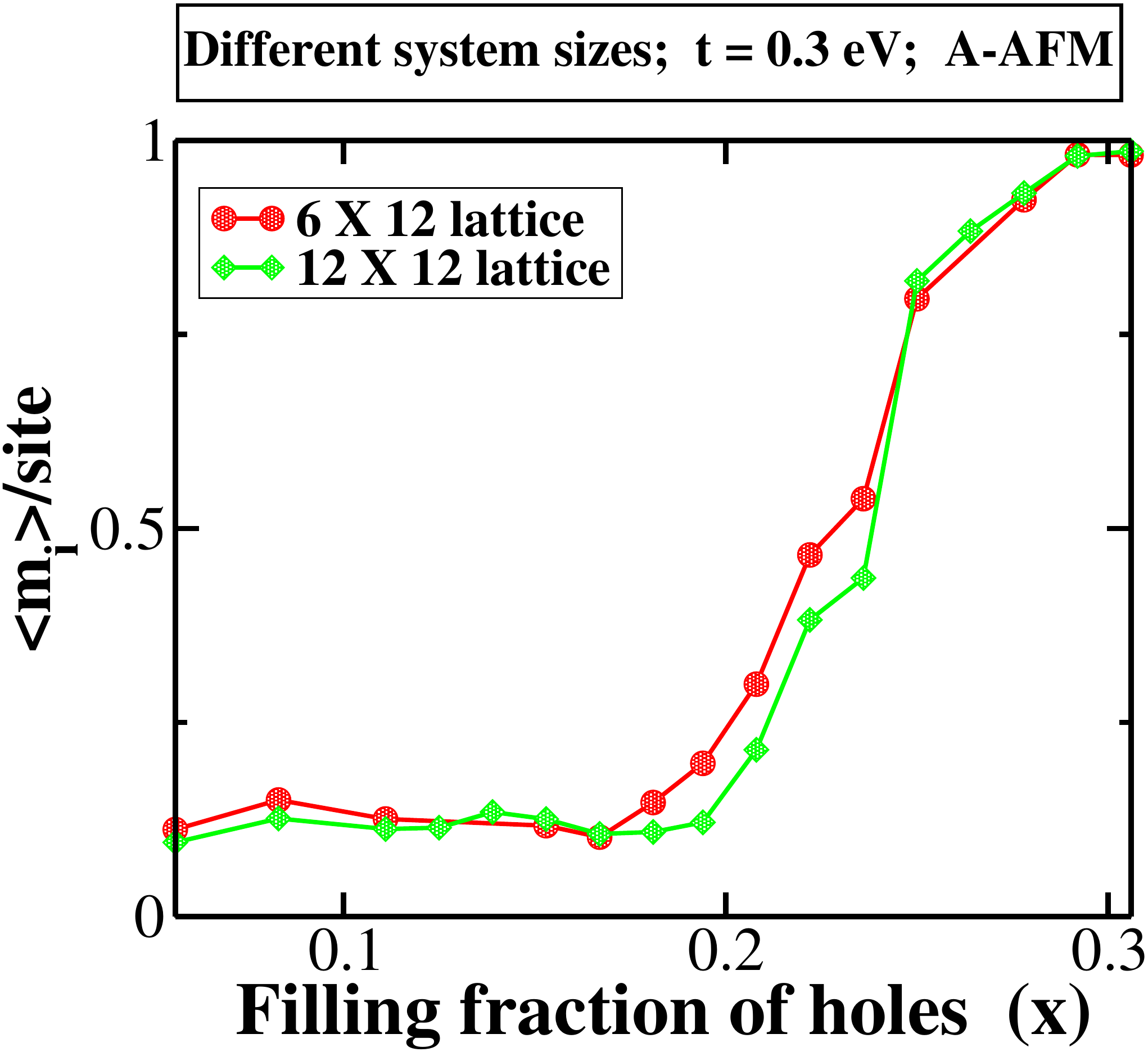}
 \caption{~(Color online)~~ Averaged~ per-site  magnetization $<m_i>$ (of 
 spins normalized to unity)
 as a function of hole doping $x$ for two different
 lattices ($6\times12$ and $12\times12$) and for a 
 fixed $T=0.001t$.
 \label{fig:sys_size_comp}
 }
\end{figure}

The hopping value $t$ is varied 
to study the 
interplay between the electron-phonon interactions 
and the superexchange interactions in the system.
The hole doping $x$ is varied as 
$0.1\leq x \leq 0.3$.
The magnetic profile still resembles that
of an A-AFM system away from the holes;
 in the NN vicinity of a hole,
the spins get polarized in the direction
of the spin of the hole thereby forming a magnetic polaron.
For  different hopping cases, the temperature variation of 
the total magnetization of the system
is studied.
We have also considered a 2D lattice with system size $12\times12$
and carried out the magnetization measurements. It shows qualitatively similar 
results as that of the $6\times 12$ lattice 
as depicted in Fig. \ref{fig:sys_size_comp}.
However,  the $12\times12$ system requires 
a running time which is $\simeq 5$ times 
that of the $6\times 12$ case; also, the number
of degenerate states for the $12\times12$ system  is much more.
Thus,
dealing with the $12 \times 12$ case
is computationally expensive. 
So, we conclude that $6\times 12$ lattice
can be considered to be representative for
the $12 \times 12$ lattice  and will be used
for investigating the
ferromagnet-insulator properties of the system.

At a very low-hole concentration $\simeq \rm 0.1$ 
(i.e., 8 particles on a
72-site lattice),
8 holes get distributed among 12 columns  such that
NN as well as NNN and NNNN interactions are avoided. In most
of the degenerate states, no two holes occupy the same column. Hence
we have site-localized holes in the system, polarizing their NN spins,
giving rise to magnetic polarons that remain  disconnected in the lattice.
Due to the NN interaction $J_{\rm xy}$, spins in a column try to align
ferromagnetically. Thus, the ferromagnetic polarons and the ferromagnetic interaction in columns together give rise to an effective low magnetization 
value {
(with a sizeable fluctuation)
}.

\begin{figure} [t]
 \includegraphics[width=4.3cm]{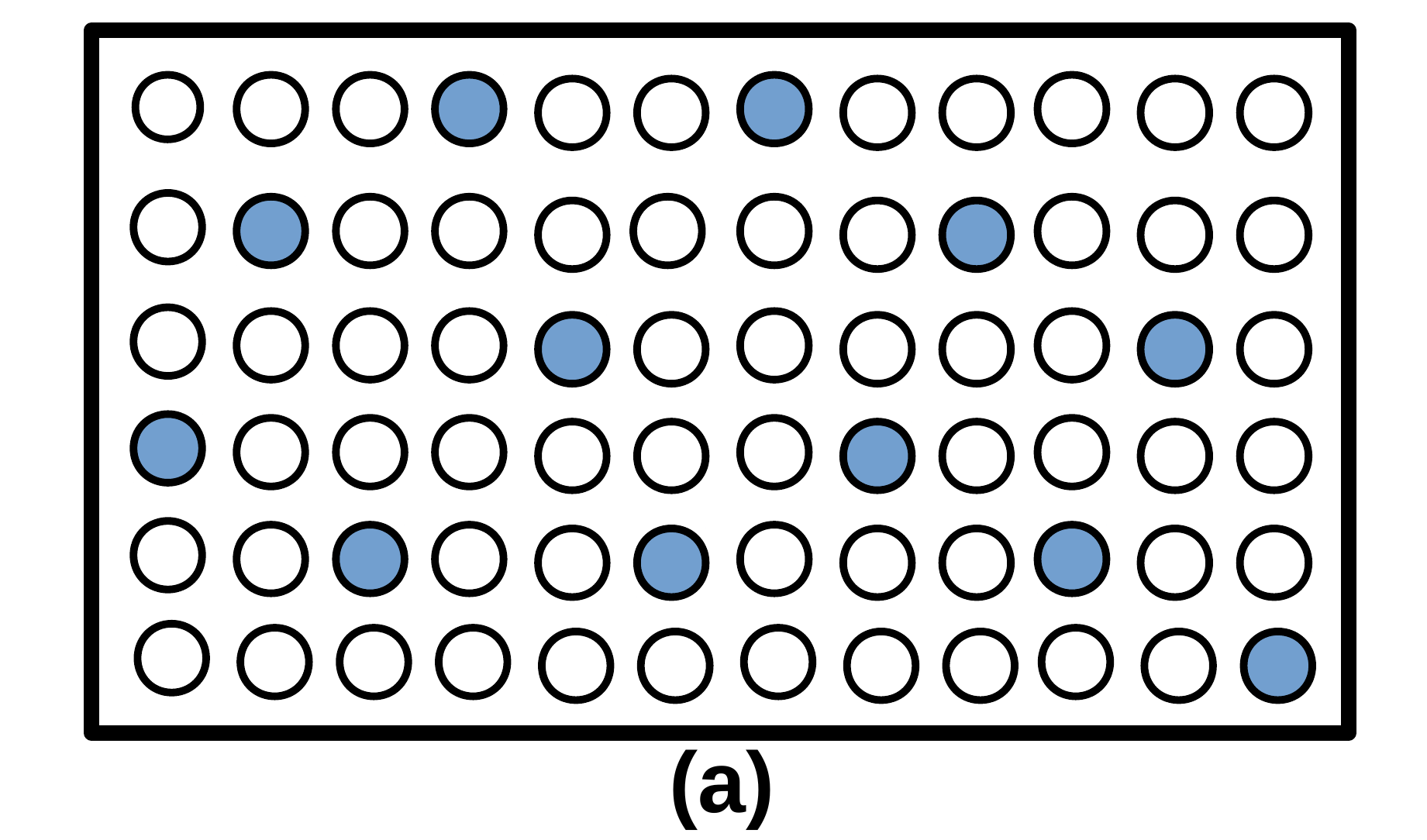}~
 \includegraphics[width=4.3cm]{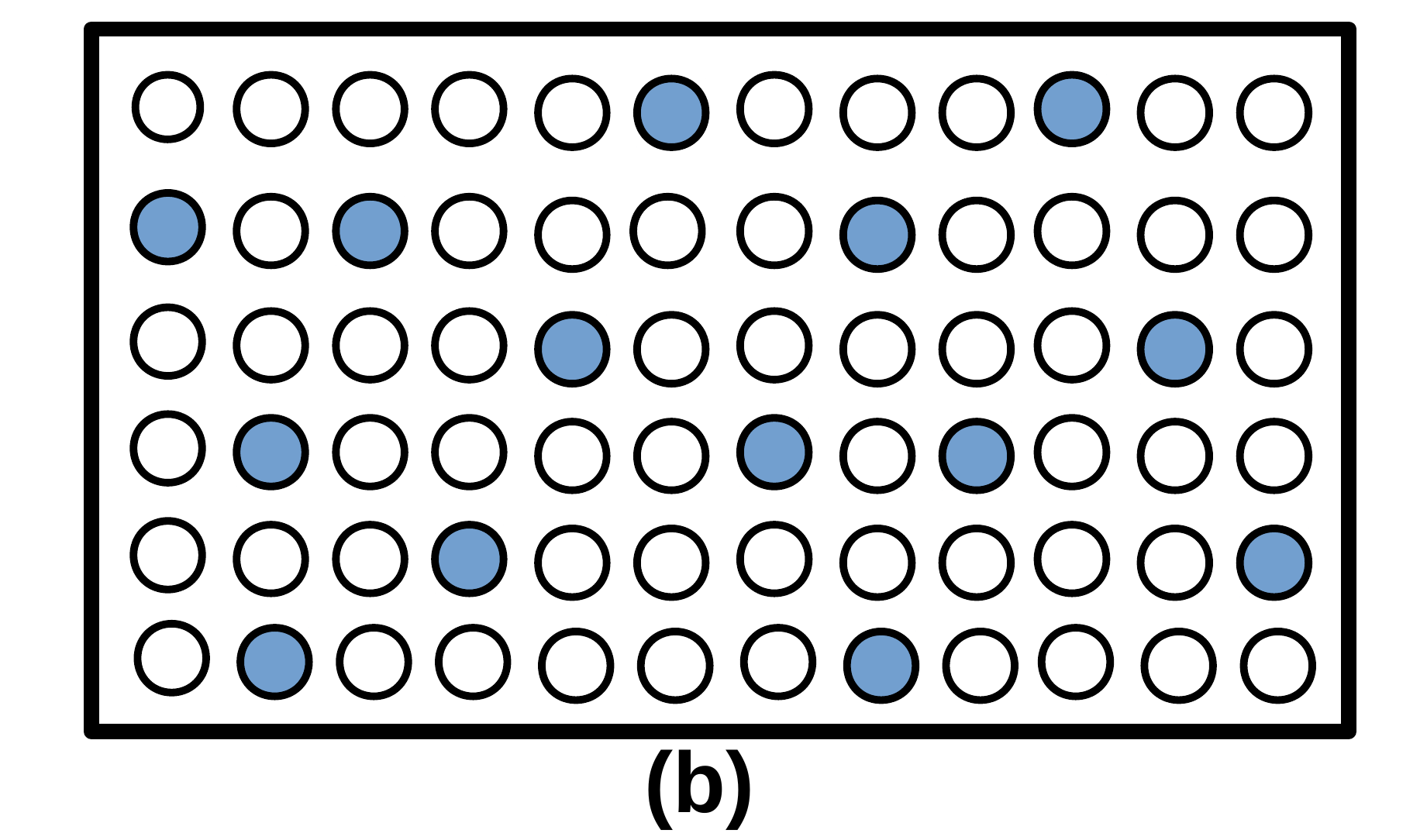}\\
 \vspace{0.2cm}
 \includegraphics[width=4.3cm]{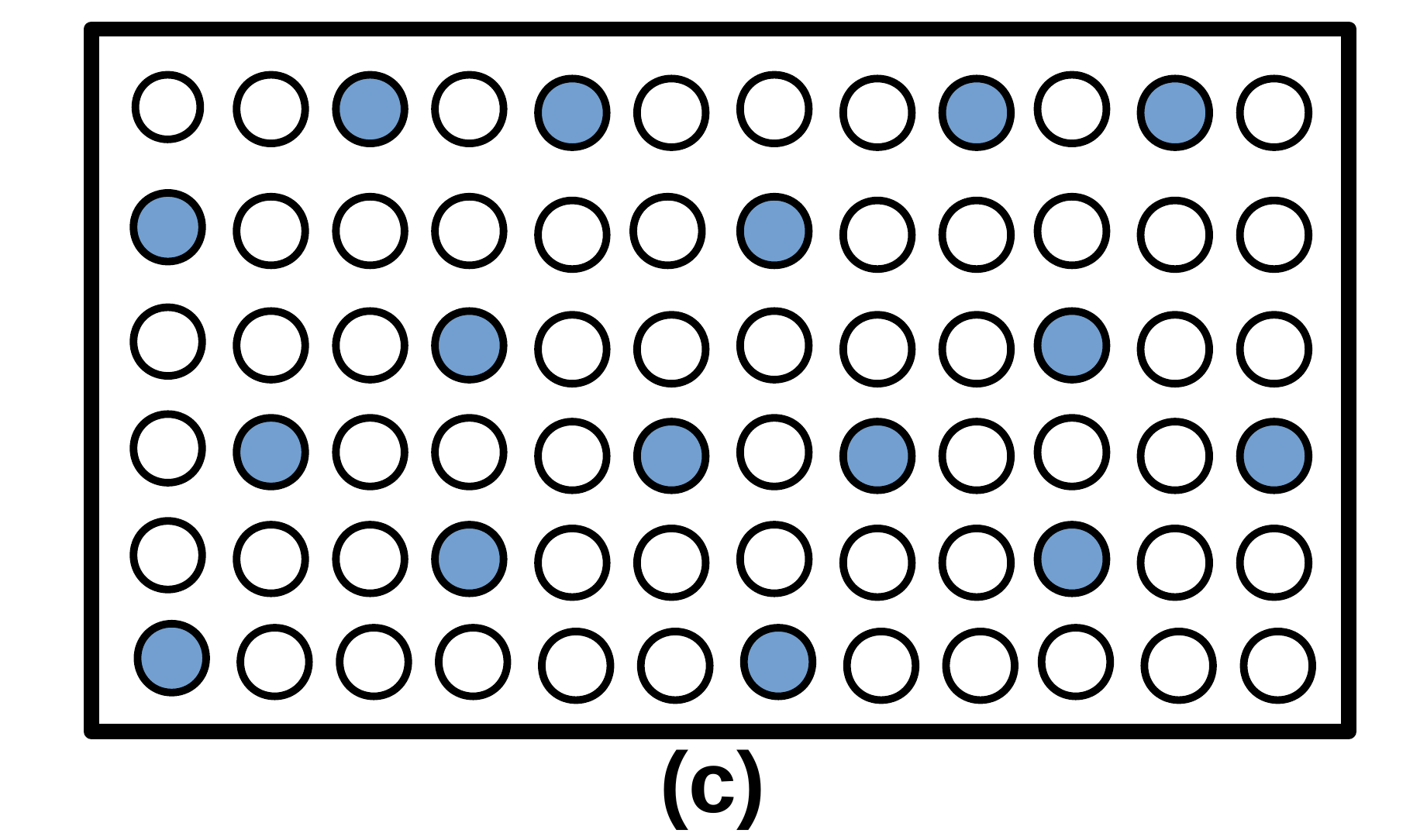}~
 \includegraphics[width=4.3cm]{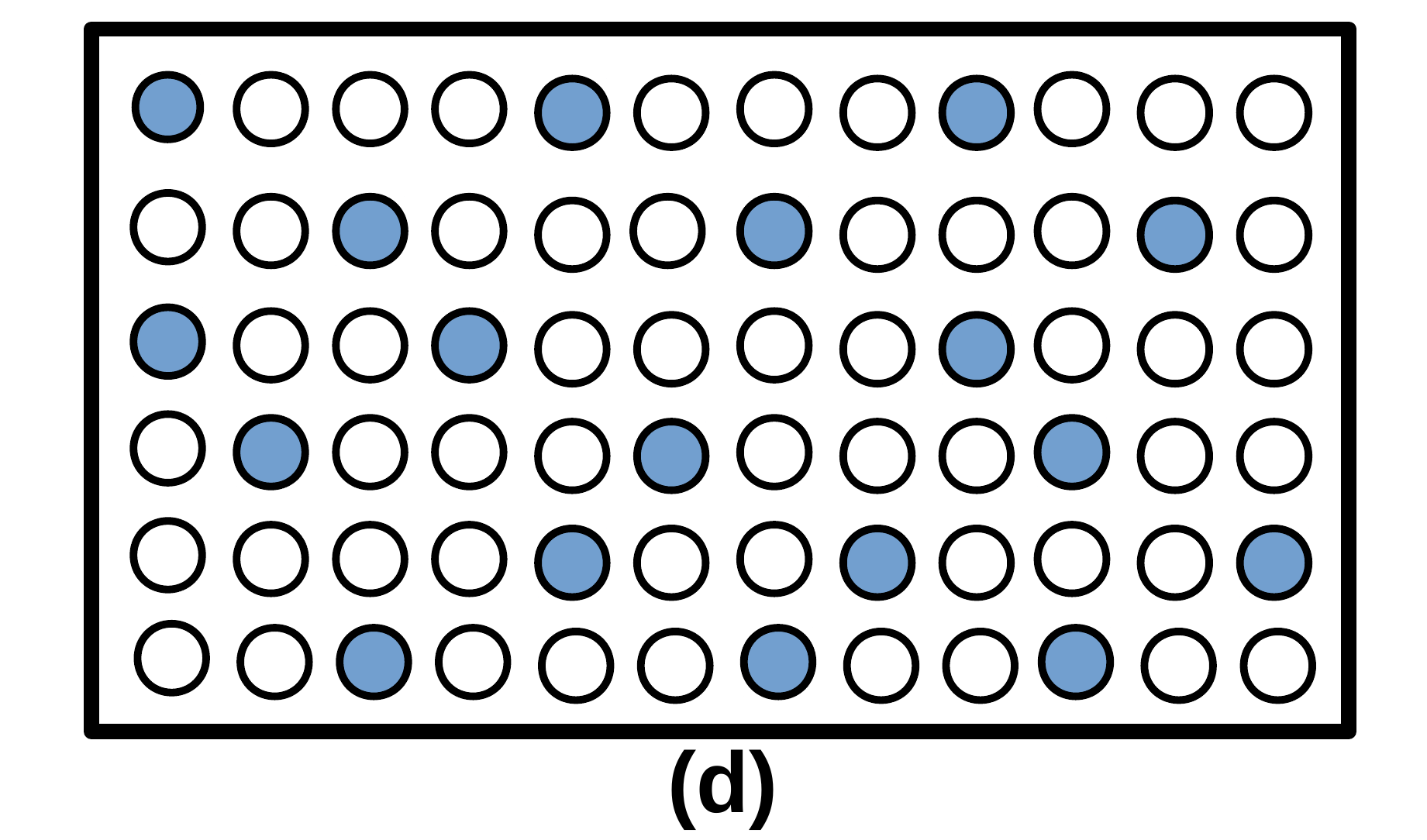}\\
 \vspace{0.2cm}
 \includegraphics[width=4.3cm]{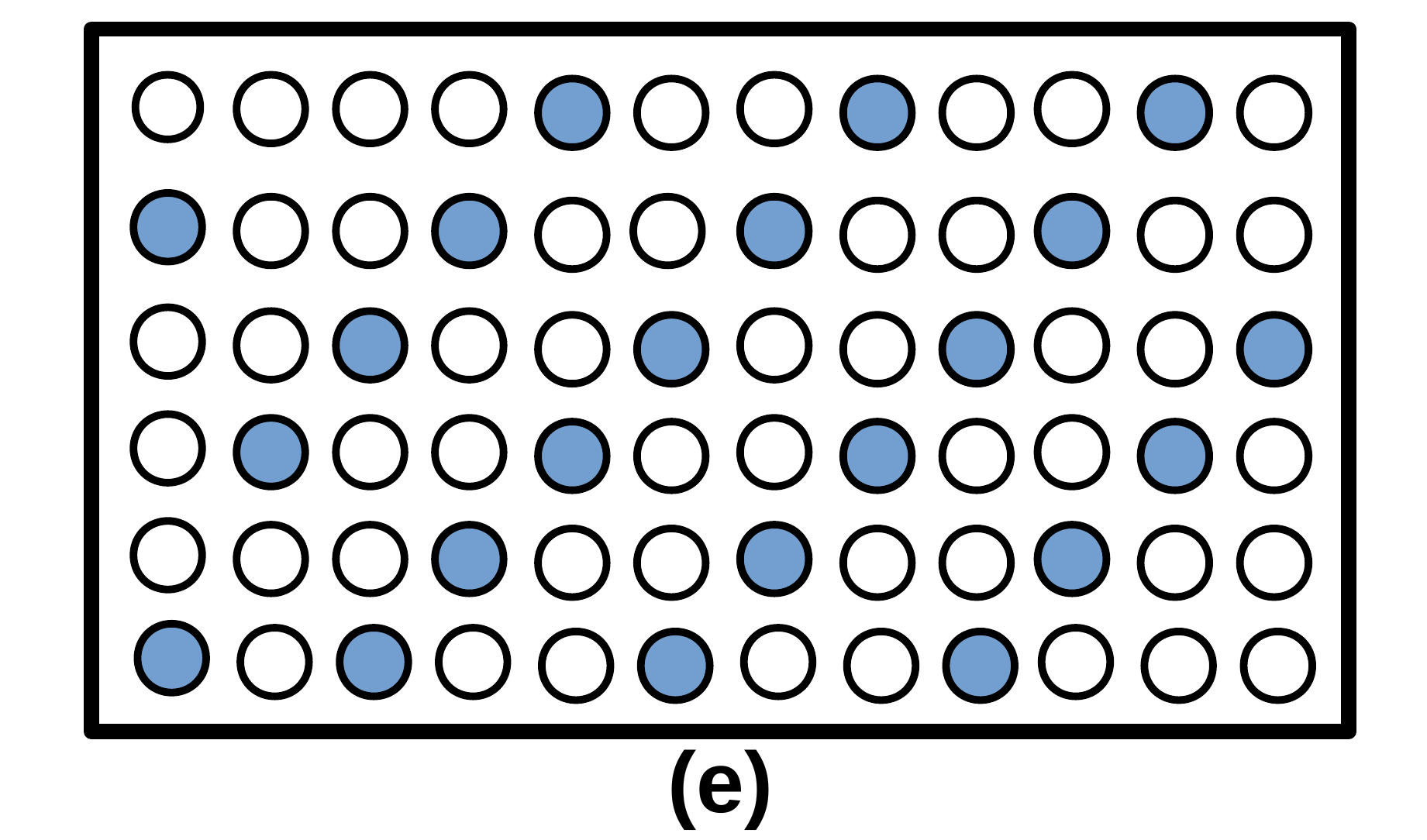}~
 \includegraphics[width=4.3cm]{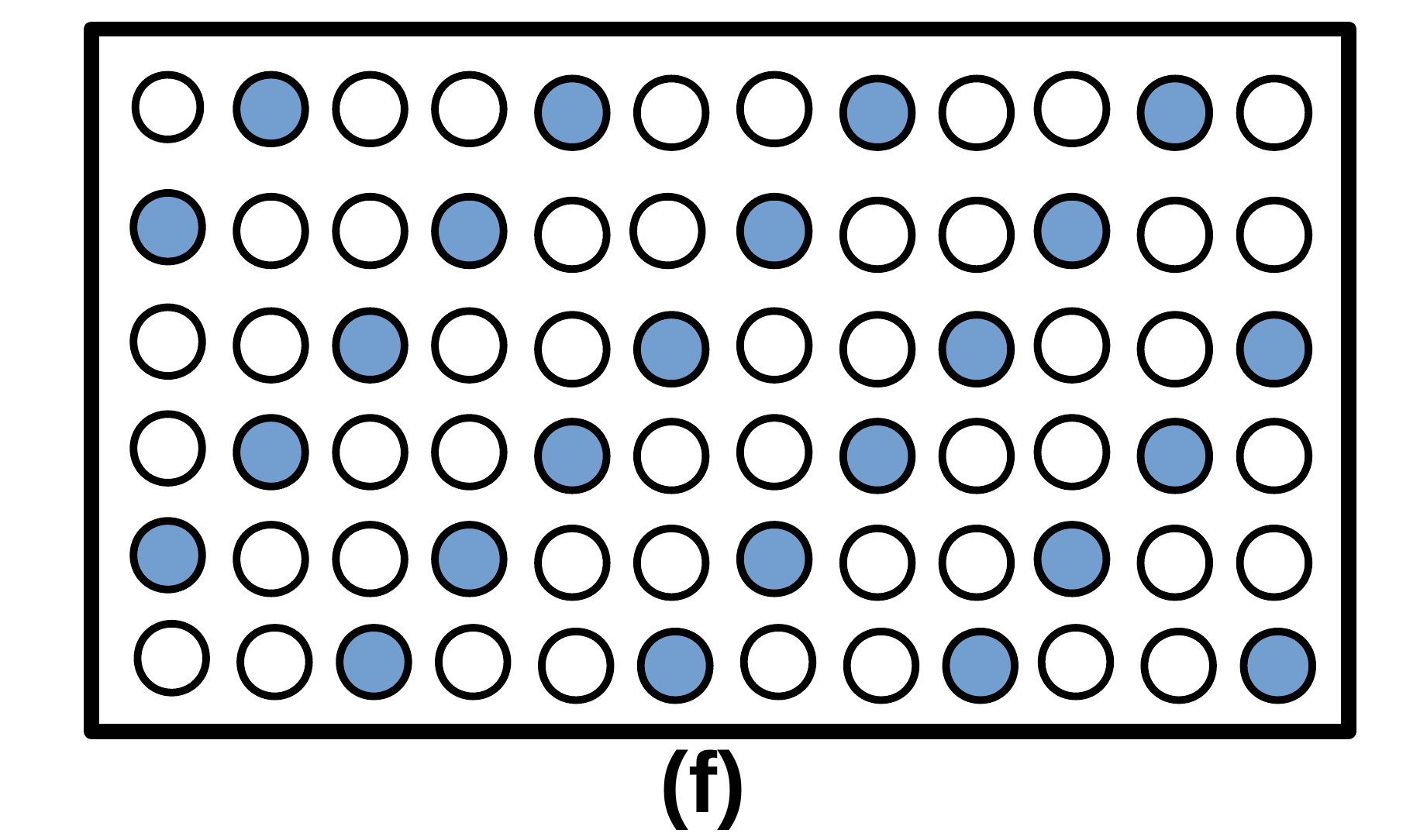}\\
 \vspace{0.2cm}
 \includegraphics[width=4.3cm]{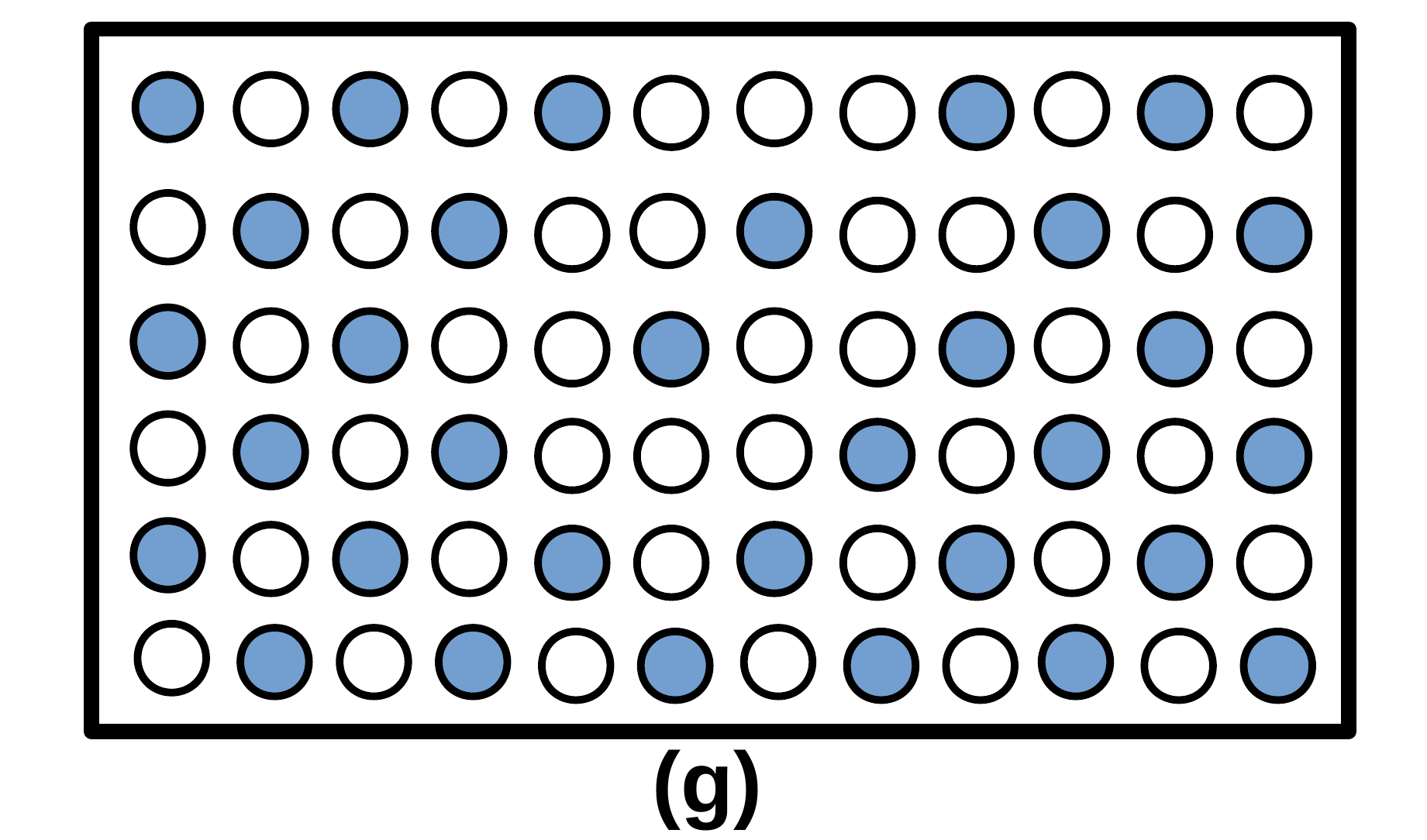}~
 \includegraphics[width=4.3cm]{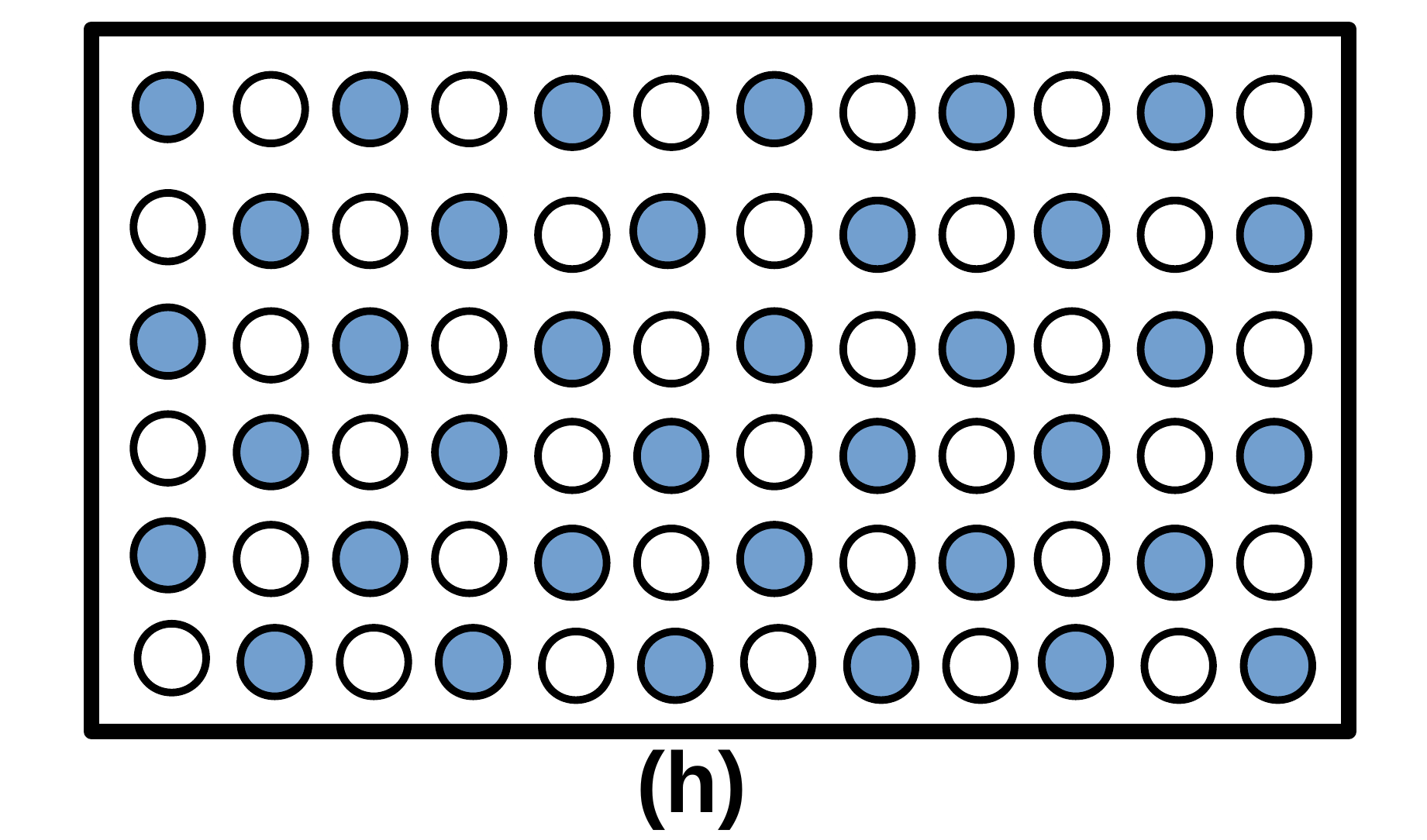}\\
 \caption{(Color online) Charge configuarations
 in the ground state of a $6\times12$ lattice. 
 An  arbitrarily
 chosen degenerate ground state, involving 72 sites, for (a) 12 holes, 
 (b) 14 holes, (c) 16 holes, 
 (d) 18 holes, (e) 22 holes,
 (f) 24 holes (diagonal
 stripe order),  (g) 32 holes, and (h) 36 holes.}
\label{fig:Charge_configs}
\end{figure}

At each temperature, the value of magnetization is essentially unchanged between hole densities $8/72$ and $12/72$ ($\simeq 0.167$);
this is because up to the filling $12/72$, holes 
can still maintain to be non-interacting (on ignoring the superexchage)
as can be seen in
Fig. \ref{fig:Charge_configs} (a).\\
At temperatures $ T < J_{\rm z}$ $(= 0.016t$ for $t=0.3$ eV and $=0.024t$ for $t=0.2$ eV),
antiferromagnetic coupling between columns is effective
and the system has low magnetization at low concentrations (i.e., $x \le 12/72$).
On increasing the temperature up to $ T 
= J_{\rm z}$, the effect of 
$J_{\rm z}$ diminishes while the effect of ferromagnetic coupling $J_{\rm xy}=1.4 J_{\rm z}$
is more dominant; thus, magnetism in the system increases with increasing temperature.
For higher temperatures $T  > J_{\rm xy}$ and again at low concentrations (i.e., $x\le 12/72$), the effect of the ferromagnetic
coupling $J_{\rm xy}$ also diminishes, and the magnetism decreases with increasing temperature.

\subsubsection{t=0.3 eV case}
\begin{figure}[t]
 \includegraphics[width=8.0cm]{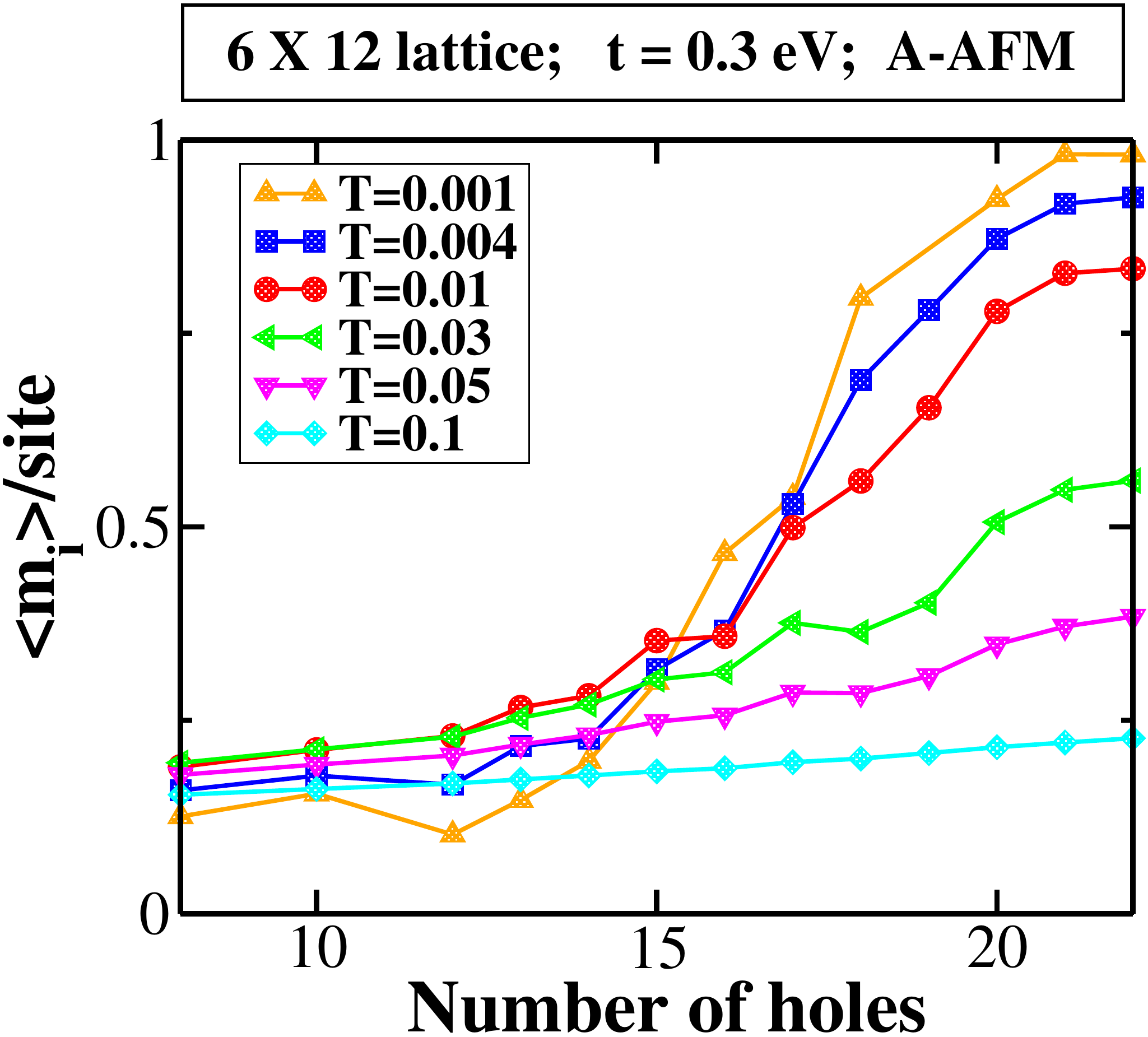}
 \caption{(Color online) Averaged per-site total magnetization $<m_i>$ (of 
 spins normalized to unity)
 as a function of the number of holes
 doped for a $6\times12$ lattice 
 and for various  temperatures (expressed in units of hopping parameter t).
 The background spin configuration is
 A-AFM type and hopping $t= 0.3$ eV.
 }
\label{fig:t=0.3_A-AFM}
\end{figure}

As the concentration of holes increases,
initially NNNN interactions and later NNN interactions become relevant;
the NN interactions being the strongest are still avoided.
Thus longer ferromagnetic chains are formed, thereby
increasing the total magnetization of the system.
So by $x= 14/72$ $\simeq 0.194$ [see Fig. \ref{fig:Charge_configs} (b)], magnetization for lower temperatures
such as $T=0.001t$ starts rising sizeably; the 
peak magnetization value is now at a reduced temperature of
$T=0.01t$.
By $x = 18/72$ $= 0.25$ [see Fig. \ref{fig:Charge_configs} (d)]
NNN interactions also appear, the different local magnetic polarons
start interacting with one another and hence form larger 
magnetic polarons and the peak magnetization
temperature  reduces
to T=0.001t; 
here, starting from $T=0.1t$, the magnetization
increases with decreasing temperature.
It is to be noted that there occurs
a narrow crossover region ($14/72 \le x \le 17/72$) where magnetization curves
for different temperatures intersect. 
In the crossover regime, there is a complex interplay of
various competing effects: 1) aligning 
of
different magnetic polarons due to dominance of $ J_{xy}$ over $ J_z$;
2) reduction of electron-hole spin interactions due to appearance of NNNN and 
NNN interactions (of strength  ${t^2\over{2E_p + 4V_p}}$);
3) commencing of percolation effects of magnetic polarons  due to large hole concentrations;
and 4) disordering effects of the temperature.
At $x = 18/72$ ($= 0.25$),
percolation effect of magnetic polarons is largely dominant
over antiferromagnetic interactions.
At even higher hole concentrations, this effect is even more pronounced;
magnetization rises faster
with lowering of temperature.
As can be seen from Fig. \ref{fig:t=0.3_A-AFM}, at lower temperatures and for $x > 15/72$,
the magnetization increases faster with increasing hole concentration.
At $T=0.001t$, we get an almost fully ferromagnetic large cluster for 
$x = 22/72$ $\simeq 0.3$  [see Fig. \ref{fig:Charge_configs} (e)], with
averaged magnetization values close to the maximum possible.

\subsubsection{t=0.2 {\rm eV} case}
\begin{figure}[t]
 \includegraphics[width=8.0cm]{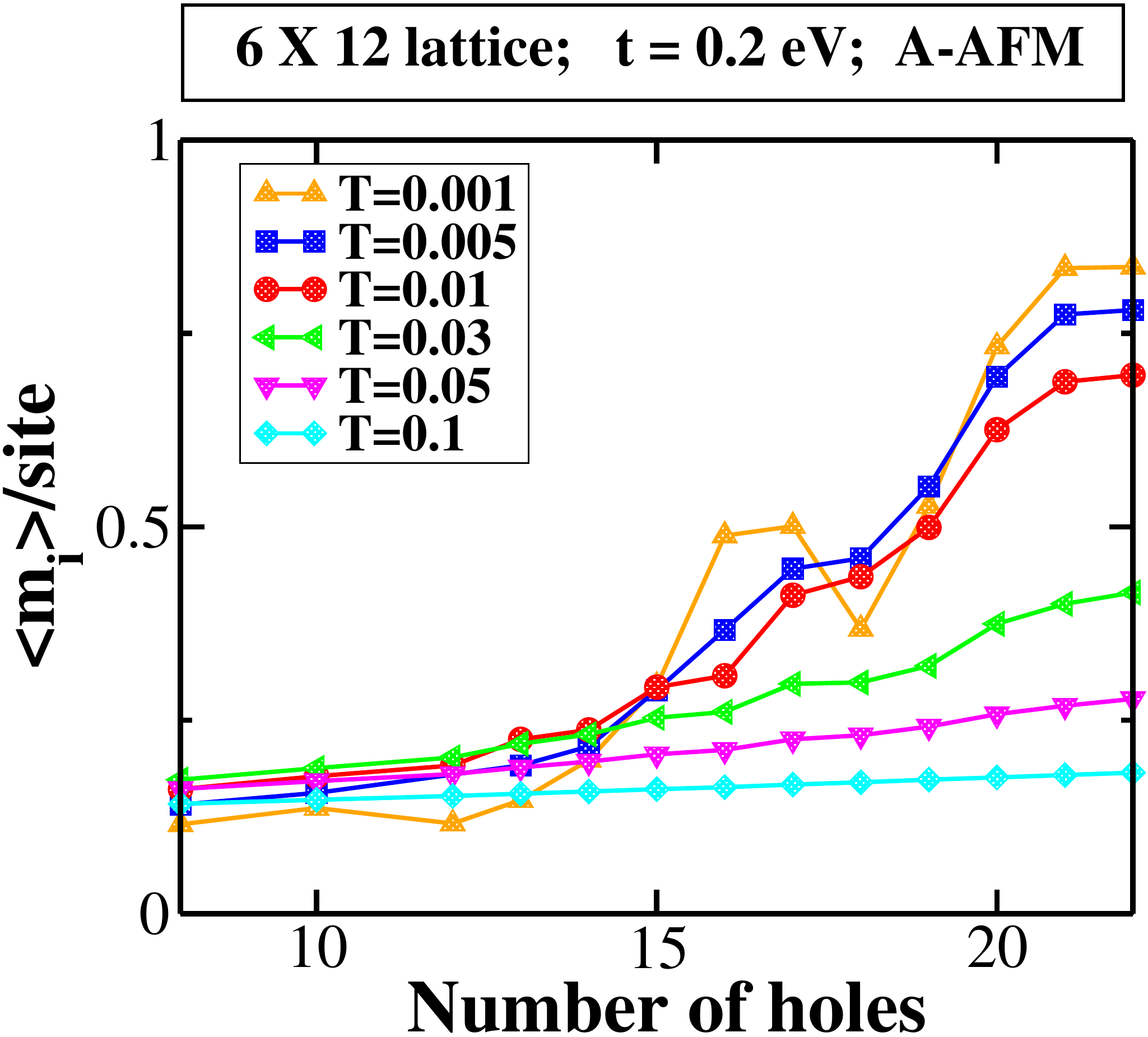}
 \caption{(Color online) Averaged per-site total magnetization $<m_i>$ (of 
 spins normalized to unity)
 as a function of the number of holes
 doped for a $6\times12$ lattice 
 and for various  temperatures (in units of hopping t).
 The background spin configuration is
 A-AFM  and $t= 0.2$ eV.
 }
\label{fig:t=0.2_A-AFM}
\end{figure}

For the  $t=0.2$  eV situation, the crossover region (i.e., $13/72 \le x \le 19/72$)
is wider than it is for the $t=0.3$  eV case.  
The peak-magnetization temperature   oscillates
in the crossover regime
(see Fig. \ref{fig:t=0.2_A-AFM}); furthermore, the curves corresponding to $T \lesssim J_z =0.024t$ intersect more than once
in the crossover region. A plausible  explanation for this
can be given as follows. The ratio of electron-hole spin interaction and antiferromagnetic coupling $\left [ \left ( {t^2\over{2E_p + 2V_p}}\right ) \Big /J_z\right ]$ 
is only 8; when NNN and NNNN interactions are relevant, the ratio reduces to $\left ({t^2\over{2E_p + 4V_p}}\right )\Big / J_z =7$.
Thus antiferromagnetic coupling becomes more prominent than for the the $t=0.3$  eV case and frustration effects 
become relevant.
It could be due to  frustration that, at lower temperatures (such as $T=0.001t$),
the magnetization curve drops  at the 
 higher carrier concentration $x = 18/72$ $= 0.25$.
 This also could be an indication
of the superspin glass phase   claimed in  experiments \cite{DD}; here,
\textquotedblleft superspin\textquotedblright refers
to a spin cluster (i.e., a large magnetic
polaron). 
At $x \ge 20/72$ $ (\simeq 0.28)$, 
percolation effect of magnetic polarons  dominates
over antiferromagnetic interactions; magnetization rises  with lowering of temperature.
Finally, for the  higher hole concentration $x = 22/72$ $\simeq 0.3$ and at $T=0.001t$, we get a 
reasonably high magnetization value of 0.85.

\subsection{G-AFM background, {\em t}=0.3 eV case}
\begin{figure}
[t]
 \includegraphics[width=8.0cm]{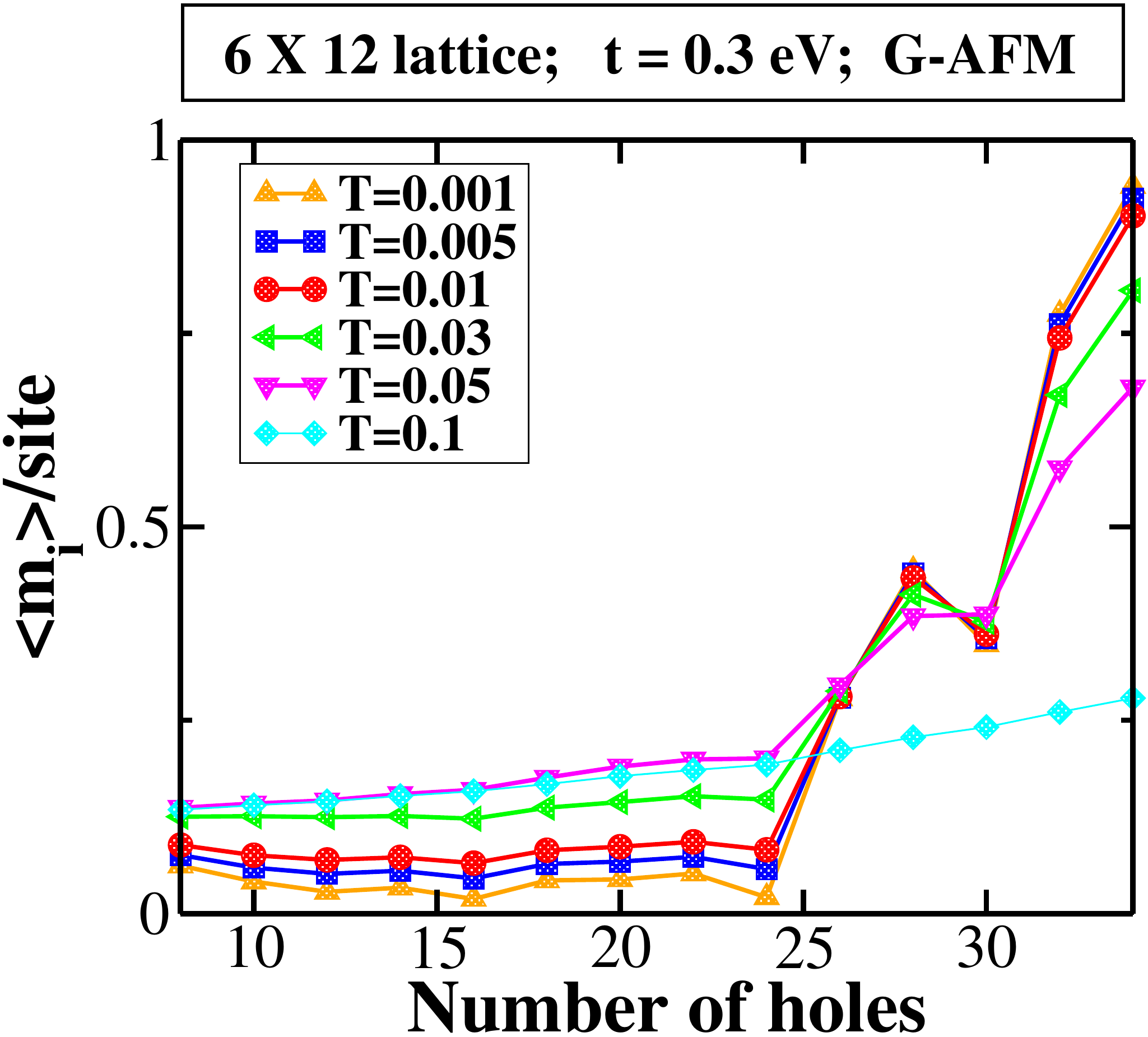}
 \caption{(Color online) Averaged per-site total magnetization $<m_i>$ (of 
 spins normalized to unity)
 as a function of the number of holes
 doped for a $6\times12$ lattice 
 and for various fixed temperatures (in units of hopping parameter t).
 The background spin configuration is
 G-AFM type and $t= 0.3$ eV.
 }
\label{fig:t=0.3_G-AFM}
\end{figure}
To gain further insight, we study the interplay between the strong ferromagnetic
electron-hole interaction that polarizes
the NN spins of a hole
and the superexchange NN antiferromagnetic interaction $J_z$.
The magnetic profile, away from the holes,
resembles that of a G-AFM system; the holes form ferromagnetic polarons
involving the hole spin and the NN spins. 
At temperatures $T \lesssim J_z =0.016t$ and hole fillings up to $x=24/72$,
due to the effect
of antiferromagnetic $ J_z$ coupling
on all sides, 
the polarizations of the magnetic polarons oppose  each other leading to a low magnetization
as shown in Fig.  \ref{fig:t=0.3_G-AFM}.
For higher temperatures, due to the dominance
of the disordering effect of the temperature over
the superexchange interaction,
there is a probability for the clusters to get 
less misaligned. Hence we notice an increase in the magnetization 
for $T > J_z$.
For $x=24/72=1/3$, we have a diagonal stripe order
as depicted in Fig. \ref{fig:Charge_configs} (f);
each column has two holes. For this 
arrangement, diagonals containing holes 
are ferromagnetic, but every such diagonal (with holes)
is antiferromagnetically coupled to its neighboring diagonal.
 In each column, half the spins are 
in one direction while the other half are  in 
the opposite direction leading to a very small total magnetization.

Here too, very similar 
to the case  of $t=0.2$ eV with A-AFM background,
there is a crossover region; the crossover occurs in the region
$24/72 < x < 32/72$.
In the crossover regime, the peak-magnetization temperature oscillates
and the curves for $ T \lesssim J z$ intersect
thrice in the crossover region.
Since all the background spins interact antiferromagnetically (which is in
contrast to the A-AFM case), percolation of magnetic polarons dominates over 
antiferromagnetic interactions 
at an even larger filling; 
 around $ x \geq 32/72$ $ = 0.44$ [refer Fig. \ref{fig:Charge_configs} (g)],
magnetization increases  with lowering of temperature..
It is to be noted that,
for half-filling [see Fig. \ref{fig:Charge_configs} (h)], we should expect a fully ferromagnetic
spin profile with a checkerboard charge structure.\cite{ME}

\subsection{Fully FM background, {\em t}=0.3 eV}
\begin{figure}[t]
 \includegraphics[width=8.0cm]{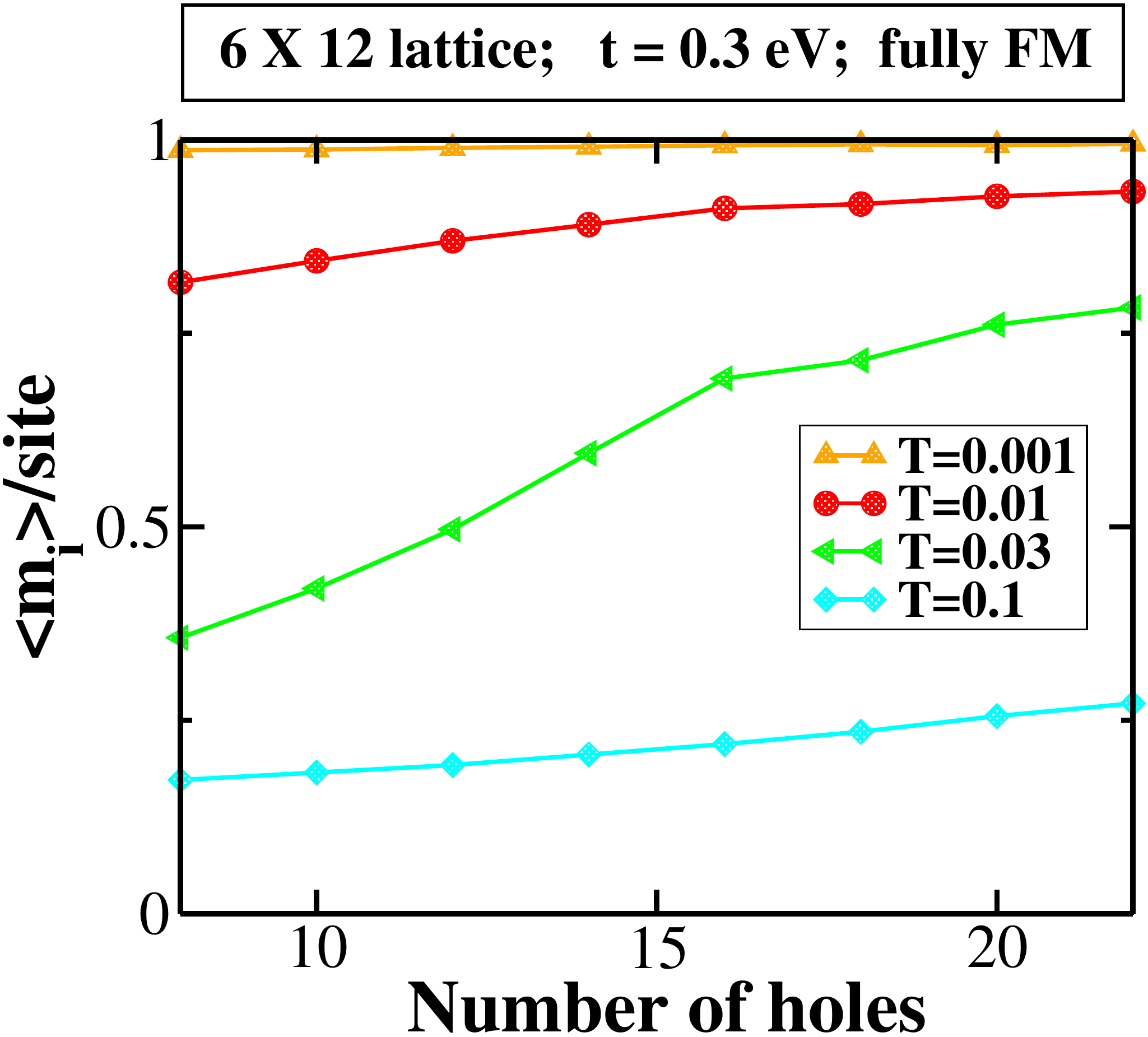}
 \caption{(Color online) Averaged per-site total magnetization $<m_i>$ (of 
 spins normalized to unity)
 as a function of the number of holes
 doped for a $6\times12$ lattice 
 and for various fixed temperatures (in units of hopping parameter t).
 The background spin configurataion is
 fully FM and $t= 0.3$ eV.
 }
\label{fig:t=0.3_FM}
\end{figure}

Lastly, to better appreciate subtleties pertaining to FMI, we also study the case
where the superexchange interaction is fully ferromagnetic (FM) with coupling $J_z = 0.016t$
when $t=0.3$ eV.
Here, while in the NN vicinity of a hole  the spins get strongly polarized thereby forming a 
ferromagnetic magnetic polaron, the magnetic profile is that of a weaker FM system away from the holes. 
Hence, for temperatures much smaller than
$ J_z$ (such as $T=0.001t$), we have an almost fully ferromagnetic system as shown in Fig. \ref{fig:t=0.3_FM},
On increasing the temperature, the spins  get more
misaligned and the magnetization reduces. At lower fillings, as 
the temperature increases to the value $0.03t$ (i.e.,  $T \approx 2 J_z$), 
the disordering effect is large enough so that the magnetization drops considerably
as shown in Fig. \ref{fig:t=0.3_FM};
On the other hand, at higher fillings and again at $T=0.03t$, percolation of magnetic polarons
counters the disordering effect and generates higher magnetization values.
For still higher temperatures (such as $T=0.1t$),
the magnetic polarons tend to orient in random directions because the superexchage coupling is ineffective,
thereby reducing the magnetization significantly.
It is interesting to note that, in all the three Figs. \ref{fig:t=0.3_A-AFM},
\ref{fig:t=0.3_G-AFM}, and \ref{fig:t=0.3_FM} plotted at $t=0.3$ eV, the magnetization curves are similar
because the superexchange is ineffective and hence the nature of superexchage coupling is irrelevant.

\section{CONCLUSIONS AND PERSPECTIVES}
We studied the nature of ferromagnetic insulator
in the experimentally relevant doping regime
of $0.1\leq x \leq 0.3$ in bulk manganites.
The magnetic interaction considered here applies
to manganites with low density of localized holes.
In regions without holes, as in the undoped manganites,
the magnetic interaction is A-AFM; in a region with 
a hole, the site-localized hole produces strong ferromagnetic coupling
between its spin and its NN electron spins.
We find that near the doping $x=0.3$, the insulator  is almost fully 
ferromagnetic. Now, the critical doping at which
the system becomes fully ferromagnetic depends on the dimension; 
in 2D it is expected to the around twice  the value
of the critical doping in three dimensions (3D) for the following reason.
In a conducting-site percolation problem, the critical concentration
for conduction in a simple cubic lattice is 0.31 and in a square lattice it is 0.59 [see Ref. \onlinecite{efros}];
hence, the critical doping to produce a percolating cluster that
is a checkerboard charge-ordered region 
is $0.5\times0.31$ in 3D and at
$0.5 \times 0.59$ in 2D [see Ref. \onlinecite{datta}].

It was experimentally observed that a FMI phase is manifested in the wide-band manganite ${\rm La_{1-x} Sr_x MnO_3}$  
in the doping region $0.1 \lesssim x \lesssim 0.18$ \cite{Book_rev}, in the intermediate-bandwidth ${\rm La_{1-x} Ca_x MnO_3}$
in the doping range $0.1 \leq x \leq 0.225$ \cite{Book_rev,LCMO_ph_diag}, and in the narrow-bandwidth 
 ${\rm Pr_{1-x} Ca_x MnO_3}$ in the region $0.1 \leq x \leq 0.3$ \cite{Book_rev}. The fact that the FMI region
 persists till a higher doping when bandwidth decreases (and concomitantly electron-phonon coupling increases \cite{peter1})
 is consistent with the fact that the tendency to localize increases as bandwidth decreases \cite{tvr,khomskii}.
 The hopping values considered in this work are pertinent to wide-bandwidth and intermediate-bandwidth manganites.
 While our one-band model (involving site-localized holes) is relevant to understand manganites 
   in the FMI region, it is certainly
 not valid to study the ferromagnet metallic (FMM) phase that occurs at higher doping in manganites;  to understand the FMM region,
 we need to invoke a two-band model
 and/or  analyze the effect of disorder on localization.
 
  The
experimental managanite phase diagram reported in Ref. \onlinecite{urushibara} reveals
increasing $T_c$ values at higher dopings  for the FMI phase  in  ${\rm La_{1-x} Sr_x MnO_3}$.
Based on this phase diagram, for a fixed $T < T_c(x=0.1)$,  we expect the magnetization to increase when  the doping increases in the FMI region; 
this is consistent with the curves in Fig. \ref{fig:t=0.3_A-AFM}. 

Lastly, comparing the $t=0.2$ eV, A-AFM case with the
$t=0.3$ eV, G-AFM case, we conclude that the antiferromagnetic coupling $J_z$
plays the  important role of 
causing frustrations in the system. 
We also point out the possible occurance
of glassiness in the system to explain the
multiple intersections of the  curves in the crossover regime at $T \lesssim J_z$ 
(see Figs. \ref{fig:t=0.2_A-AFM} and \ref{fig:t=0.3_G-AFM}).
Such a picture is supported by the observed  superspin glass
phase in ${\rm La _{0.82} Ca_{0.18} MnO_3}$ ferromagnetic insulator
at $T \lesssim 70$ K \cite{DD}.
Further theoretical analysis is 
required to clearly identify and characterize a superspin glass phase
at lower temperatures.

\section{acknowledgements} 
The authors would like to thank A. Ghosh and P. B. Littlewood for useful discussions.


\begin{thebibliography}{1}
\bibitem{cnr}
{\em Colossal Magnetoresistance, Charge Ordering, and Related
Properties of Manganese Oxides}, edited by C.N.R. Rao and
B. Raveau (World Scientific, Singapore, 1998).
\bibitem{khomskii0}
D. I. Khomskii, Physica Scripta 72, CC8-14 (2005).
\bibitem{hotta}
T. Hotta, Rep. Prog. Phys. 69, 2061 (2006).
\bibitem{tokura}
Y. Tokura, Rep. Prog. Phys. {\bf 69}, 797 (2006).
\bibitem{cheong}
See K.H. Kim, M. Uehara, V. Kiryukhin and S.-W. Cheong, in 
{\em Colossal Magnetoresistive Manganites}, edited by T. Chatterji, (Kluwer Academic, Dordrecht, 2004).
\bibitem{raveau}
 C. Martin, A. Maignan, M. Hervieu, and B. Raveau,
 Phys. Rev. B {\bf 60}, 12191 (1999).
\bibitem{littlewood}
A. J. Millis, P. B. Littlewood, and B. I. Shraiman, Phys.
Rev. Lett. {\bf 74}, 5144 (1995).
\bibitem{pinaki}
Sanjeev Kumar and Pinaki Majumdar,
Phys. Rev. Lett. {\bf 96}, 016602 (2006);  S. Kumar and P. Majumdar, cond-mat 0406082.
\bibitem{cbm2d}
A. Ghosh and S. Yarlagadda,
Phys. Rev. B {\bf 96}, 125108 (2017).
\bibitem{dai1}
P. C. Dai, J. A. Fernandez-Baca, E. W. Plummer, 
Y. Tomioka, and Y. Tokura,
Phys. Rev. B {\bf 64}, 224429 (2001).
\bibitem{dai2}
P. C. Dai, J. A. Fernandez-Baca, N. Wakabayashi,
E. W. Plummer, Y. Tomioka, and Y. Tokura,
Phys. Rev. Lett. {\bf 85}, 2553 (2000).
\bibitem{jiang}
W. Jiang, X. Z. Zhou, G. Williams, R. Privezentsev, and
Y. Mukovskii,  Phys. Rev. B {\bf 79}, 214433 (2009).
\bibitem{markovich1}
V. Markovich, E. Rozenberg, A. I. Shames, G. Gorodetsky,
I. Fita, K. Suzuki, R. Puzniak, D. A. Shulyatev, and
Y. M. Mukovskii,  Phys. Rev. B {\bf 65}, 144402 (2002).
\bibitem{markovich2}
V. Markovich, I. Fita, R. Puzniak, M. I. Tsindlekht,
A. Wisniewski, and G. Gorodetsky, Phys. Rev. B {\bf 66}, 094409 (2002).
\bibitem{peter1}
T. F. Seman, K. H. Ahn, T. Lookman, A. Saxena, A. R. Bishop, and P. B. Littlewood,
Phys. Rev. B {\bf 86}, 184106 (2012).
\bibitem{peter2}
G. G. Guzm\'an-Verri, R. T. Brierley, P. B. Littlewood, e-print  arXiv:1701.02318.
\bibitem{tvr}
G. V. Pai, S. R.  Hassan, H. R. Krishnamurthy, and
T. V. Ramakrishnan,   Europhys. Lett. {\bf 64}, 696 (2003);
T. V. Ramakrishnan,  H. R. Krishnamurthy, S. R.  Hassan, and G. V. Pai, Phys. Rev. Lett. {\bf 92}, 157203 (2004).
\bibitem{khomskii}
M. Yu. Kagan, D. I. Khomskii, and M.V. Mostovoy, Eur. Phys. J. B {\bf 12}, 217 (1999).
\bibitem{DD}
P. A. Kumar, R. Mathieu, P. Nordblad, S. Ray, O. Karis, G. Andersson, 
and D. D. Sarma, Phys. Rev. X {\bf 4}, 011037 (2014).
\bibitem{ME}
S. Paul, R. Pankaj, S. Yarlagadda, P.  Majumdar, and P. B. Littlewood,
Phys. Rev. B {\bf 96}, 195130 (2017).
\bibitem{Book_rev}
V. Markovich, A. Wisniewski, and H. Szymczak, Magnetic 
properties of perovskite manganites and their modifications, in  
{\em{Handbook of Magnetic Materials}} 
 {\bf 22},   1-201 (2014), 
 edited by K. H. J. Buschow  
(New York:  Elsevier). 
\bibitem{LCMO_ph_diag}
P. Schiffer, A. P. Ramirez, W. Bao, and S.-W. Cheong, Phys. Rev.
Lett. 75, 3336 (1995).
\bibitem{super_ex}
K. Hirota, N. Kaneko, A. Nishizawa, and Y. Endoh, 
J. Phys. Soc. Jpn. {\bf 65}, 3736 (1996͒); F. Moussa, M. Hennion, 
J. Rodr\'{i}guez-Carvajal, H. Moudden, L. Pinsard, and A. Revcolevschi,
Phys. Rev. B {\bf 54}, 15149 (1996͒); G. Biotteau, M. Hennion, F. Moussa,
J. Rodr\'{i}guez-Carvajal, L. Pinsard, A. Revcolevschi,
Y. M. Mukovskii, and D. Shulyatev, {\em ibid}. {\bf 64}, 104421 (2001).
\bibitem{khali}
N. N. Kovaleva, Andrzej M. Ole\'s, A. M. Balbashov, A. Maljuk,
D. N. Argyriou, G. Khaliullin, and B. Keimer,
Phys. Rev. B {\bf 81}, 235130 (2010).
\bibitem{gennes}
P. -G. de Gennes, Phys. Rev. {\bf 118}, 141 (1960).
\bibitem{Izyumov}
Y. A. Izyumov, Y. N. Skryabin, Phys.-Usp. {\bf 44}, 109 (2001).
\bibitem{anderson}
P. W. Anderson, Phys. Rev. {\bf 115},  2 (1959).
\bibitem{rpsy}
R. Pankaj and S. Yarlagadda, 
Phys. Rev. B {\bf 86}, 035453 (2012)
\bibitem {A_Dey} A. Dey, M. Q. Lone, 
and S. Yarlagadda, Phys. Rev. B {\bf 92}, 094302
(2015).
\bibitem{efros}
A. L. Efros, {\em Physics and Geometry of Disorder}  (Mir
Publishers, Moscow, 1986).
\bibitem{datta}
S. Datta and S. Yarlagadda, Proceedings of the DAE Solid State
Physics Symposium (Golden
Jubilee) {\bf 50}, 605 (2005).
\bibitem{urushibara}
A. Urushibara, Y. Moritomo, T. Arima, A. Asamitsu, G. Kido, and Y. Tokura
Phys. Rev. B {\bf 51}, 14103 (1995).


\end{thebibliography}
\end{document}